\documentclass{emulateapj}

\newcommand{\deluxetablestar}{deluxetable*} 

\newcommand{\mc}{\multicolumn{2}{c}}
\newcommand{\mcf}{\multicolumn{4}{c}}

\newcommand{\Ha}{\ensuremath{\mathrm{H} \alpha}}

\newcommand{\sii}{[\ion{S}{2}]}
\newcommand{\nii}{[\ion{N}{2}]}
\newcommand{\DM}{\ensuremath{\mathrm{DM}}}
\newcommand{\EM}{\ensuremath{\mathrm{EM}}}

\newcommand{\Ms}{\ensuremath{\mathcal{M}_s}}
\newcommand{\Ma}{\ensuremath{\mathcal{M}_A}}
\newcommand{\emsinb}{\ensuremath{\EM \sin |b|}}

\newcommand{\cucm}{\ensuremath{\textrm{ cm}^{-3}}}
\newcommand{\cmsix}{\ensuremath{\textrm{ cm}^{-6}}}
\newcommand{\kms}{\ensuremath{\textrm{ km s}^{-1}}}
\newcommand{\kpc}{\ensuremath{\textrm{ kpc}}}
\newcommand{\pc}{\ensuremath{\textrm{ pc}}}
\newcommand{\K}{\ensuremath{\textrm{ K}}}

\newcommand{\Pini}{\ensuremath{p_\mathrm{ini}}}
\newcommand{\Bext}{\ensuremath{B_\mathrm{ext}}}
\newcommand{\obs}{\ensuremath{\mathrm{obs}}}
\newcommand{\ini}{\ensuremath{\mathrm{ini}}}

\newcommand{\hwim}{\ensuremath{h_\mathrm{WIM}}}
\newcommand{\hsimbox}{\ensuremath{L_\mathrm{box}}}
\newcommand{\vlsr}{\ensuremath{V_\mathrm{LSR}}}

\begin{document}


\submitted{ApJ, in press}

\title{The Turbulent Warm Ionized Medium: Emission Measure Distribution and MHD Simulations}
\author{Alex S. Hill}
\affil{Department of Astronomy, University of Wisconsin-Madison, Madison, WI 53706}
\email{hill@astro.wisc.edu}
\author{Robert A. Benjamin}
\affil{Department of Physics, University of Wisconsin-Whitewater, Whitewater, WI 53190}
\author{Grzegorz Kowal, Ronald J. Reynolds, L. Matthew Haffner, and Alex Lazarian}
\affil{Department of Astronomy, University of Wisconsin-Madison, Madison, WI 53706}

\begin{abstract}
We present an analysis of the distribution of \Ha\ emission measures for the warm ionized medium (WIM) of the Galaxy using data from the Wisconsin H-Alpha Mapper (WHAM) Northern Sky Survey. Our sample is restricted to Galactic latitudes $|b| > 10 \arcdeg$. We removed sightlines intersecting nineteen high-latititude classical \ion{H}{2} regions, leaving only sightlines that sample the diffuse WIM. The distribution of \emsinb\ for the full sample is poorly characterized by a single normal distribution, but is extraordinarily well fit by a lognormal distribution, with $\langle \log \emsinb (\pc \cmsix)^{-1} \rangle = 0.146 \pm 0.001$ and standard deviation $\sigma_{\log \emsinb} = 0.190 \pm 0.001$.  $\langle \log \emsinb \rangle$ drops from $0.260 \pm 0.002$ at Galactic latitude $10< |b|< 30$ to $0.038 \pm 0.002$ at Galactic latitude $60< |b|< 90$. The distribution may widen slightly at low Galactic latitude. We compare the observed EM distribution function to the predictions of three-dimensional magnetohydrodynamic simulations of isothermal turbulence within a non-stratified interstellar medium. We find that the distribution of \emsinb\ is well described by models of mildy supersonic turbulence with a sonic Mach number of $\sim 1.4 - 2.4$. The distribution is weakly sensitive to the magnetic field strength. The model also successfully predicts the distribution of dispersion measures of pulsars and \Ha\ line profiles. In the best fitting model, the turbulent WIM occupies a vertical path length of $400-500 \pc$ within the $1.0-1.8 \kpc$ scale height of the layer. The WIM gas has a lognormal distribution of densities with a most probable electron density $n_{pk} \approx 0.03 \cucm$. We also discuss the implications of these results for interpreting the filling factor, the power requirement, and the magnetic field of the WIM.
\end{abstract}


\section{Introduction}

The warm ionized medium (WIM, sometimes called the diffuse ionized gas, or DIG) is a major component of our Galaxy, consisting of a pervasive, diffuse plasma layer with temperatures near $8000$~K, a scale height of about $1 \kpc$ \citep[more than three times the scale height of the neutral hydrogen;][]{r89a}, and a space-averaged midplane density of $0.03 \cucm$ \citep{f01}. By combining pulsar dispersion measures with \Ha\ emission measures, \citet{r91} characterized the gas with a simple model consisting of discrete clumps with densities of $0.08 \cucm$ occupying $\sim 20 - 40 \%$ of the volume of a $2 \kpc$ thick, plane parallel layer about the midplane of the Galaxy. The WIM is distinct from a component of warm, ionized gas associated primarily with classical \ion{H}{2} regions, which has a scale height of $40 - 70 \pc$ and a space-averaged density comparable to that of the WIM \citep{f01, gbc01}. Thus, in directions away from the Galactic midplane, emission from warm, ionized interstellar gas is dominated by the WIM.

The ionization power requirement of the WIM is $2 n^2 \alpha^{(2)} f \hwim \sim 4 \times 10^6$ ionizing photons $\textrm{ s}^{-1} \textrm{ cm}^{-2}$ or $\sim 2 \times 10^{-4} \textrm{ erg s}^{-1} \textrm{ cm}^{-2}$ in the Galactic disk \citep{r90b}, where $\alpha^{(2)}$ is the recombination coefficient to levels $\ge 2$, and $n$, $f$ and $\hwim$ are the electron density, volume filling fraction, and scale height of the WIM, respectively. With a core collapse supernova rate of one $10^{51}$~erg event per $50$ years \citep{dhk06} in the Galactic disk ($r \sim 15 \kpc$), supernovae inject barely enough energy to ionize the WIM even if all of the energy from supernovae goes into ionizing the WIM. Lyman continuum radiation from O stars injects $3 \times 10^7 \textrm{ ionizing photons s}^{-1} \textrm{ cm}^{-2}$ \citep{a82}; this is the only known source that injects ample energy to ionize the WIM. With the O stars and supernovae concentrated in the plane, the mechanism by which the ionizing radiation and shocks reach the $1 \kpc$ scale height is not well understood.

Given the observed velocities and expected sources of viscosity, one expects the WIM to have a high Reynolds number \citep{b99}, meaning the gas should be turbulent. At small scales, observations have confirmed this expectation. \citet{ars95} used radio measurements of interstellar scintillation and fluctuations in dispersion measure and rotation measure to demonstrate that the ISM is a turbulent medium with density fluctuations following a Kolmogorov-like power law over scales from $2 \times 10^8$~cm ($10^{-5}$~AU) to at least $10^{15}$~cm ($70$~AU). Their data are consistent with a power law spectrum to scales as large as $10^{20}$~cm ($30 \pc$).

Astrophysical turbulence does not need to be Kolmogorov-like, however. (See review by \citealt{l06}.) Integrated intensity fluctuations of $^{13}$CO and $^{12}$CO show a shallow spectrum \citep{sbh98}. Using models of radiative transfer in a turbulent medium, \citet{lp04} interpreted these fluctuations as corresponding to a spectrum of turbulent density $E(k)\sim k^{-\beta}$ with a spectral index $\beta=0.8$, whereas $\beta=5/3$ for Kolmogorov turbulence. \citet{sw08} found $\beta=1-2$ for fluctuations in the optically thin isotope C$^{18}$O in the low mass, star-forming molecular cloud Lynds 1551. \citet{cls06} showed that \ion{H}{1} data for high latitudes correspond to a velocity spectrum steeper than the Kolmogorov value and a density spectrum shallower than Kolmogorov. This is well in agreement with magnetohydrodynamic (MHD) numerical simulations \citep{blc05} which show that supersonic turbulence tends to have steep velocity and shallow density spectra. \citet{kr05} found similar shallow density spectra at high Mach number in hydrodynamic simulations and estimated a sonic Mach number of order unity for the WIM.

Pulsar dispersion measures provide the column density through the gas, which \citet{mt81}, \citet{tc93}, and \citet{gbc01}, among others, used to construct models of the distribution of warm, ionized gas in the Galaxy. The bulk of the information about the temperature, ionization state, and kinematics of the WIM has come from the study of faint, optical emission lines such as \Ha\ ($6563 \mathrm{\AA}$), \sii\ $\lambda 6716$, and \nii\ $\lambda6583$ with Fabry-Perot spectrometers such as the Wisconsin H-Alpha Mapper (WHAM). Velocity-resolved observations of \Ha\ emission reveal the distribution and kinematics of the gas, while the line ratios \nii$/$\Ha\ and \sii$/$\nii\ are sensitive primarily to temperature and ionization state, respectively \citep{hrt99}. The comparison of \Ha\ emission measure with pulsar dispersion measures along the same sightline allow estimates of the ``clumpiness'' of the gas \citep{r91}.

In this paper, we show that the \Ha\ emission measure of the WIM is extremely well characterized by a lognormal distribution, demonstrate that the observed distribution is plausibly produced by MHD turbulence in an approximately isothermal medium, and pursue observational and theoretical ramifications of this result. In \S~\ref{sec:data}, we discuss the WHAM data used in this work and our masking of the WHAM data to study only the diffuse WIM. In \S~\ref{sec:dist}, we demonstrate that the diffuse WIM has a lognormal distribution of emission measure, and we discuss the occurrence of lognormal distributions in connection with turbulence in \S~\ref{sec:turbulence}. We compare the results of isothermal MHD models of turbulence with the data in  \S~\ref{sec:model}, finding a reasonable correpsondance.  In \S~\ref{sec:tests}, we use these models to make and then test additional predictions for the WIM  observations. Finally, in \S~\ref{sec:implications}, we explore the ramifications of this work on the filling fraction, power requirement, and magnetic field structure of the WIM.

\section{WHAM data} \label{sec:data}

\begin{deluxetable}{l c r@{.}l r@{.}l r}

\tabletypesize{\footnotesize}
\tablecolumns{7}
\tablewidth{0pt}
\tablecaption{Large Classical \ion{H}{2} regions, $\delta > -30 \arcdeg$, $|b| > 10 \arcdeg$}

\tablehead{
    \multicolumn{6}{c}{Ionizing Star} & \colhead{Radius} \\
    \cline{1-6}
    \colhead{} & \colhead{Spectral} & \mcf{} & \colhead{Removed} \\
    \colhead{Name} & \colhead{Type} & \mc{$l$ (\arcdeg)} & \mc{$b$ (\arcdeg)} & \colhead{(\arcdeg)}
}

\startdata
\cutinhead{Classical \ion{H}{2} regions}
$\zeta$ Oph     & O9 V           &  6&3 &  23&6 & 7 \\ 
HD 161056       & B1.5 V         & 18&7 &  11&6 & 2 \\ 
$\pi$ Aqr       & B1 Ve          & 66&0 & -44&7 & 3 \\ 
10 Lac          & O9 V           & 96&6 & -17&0 & 4 \\ 
AO Cas          & O9 IIInn+...   &117&6 & -11&1 & 3 \\ 
PHL 6783\tablenotemark{a} & sd:B &123&6 &-73&5 & 3 \\
$\phi$ Per      & B2 Vpe         &131&3 & -11&3 & 8 \\ 
$\alpha$ Cam    & O9.5 Iae       &144&1 &  14&0 & 6 \\ 
$\xi$ Per       & O7.5 IIIe      &160&4 & -13&1 & 10 \\ 
HD 41161        & O8 V           &165&0 &  12&9 & 5 \\ 
$\lambda$ Ori (HD 36861) & O8 III &195&0 &-12&0 & 7 \\ 
$\sigma$ Ori    & O9.5 V         &206&8 & -17&3 & 9 \\
PG 1034+001     & DO            &247&6 &  47&8 & 5 \\
PG 1047+003     & sd:B          &250&9 &  50&2 & 5 \\ 
Spica ($\alpha$ Vir) & B1 III-IV+... &316&1 & 50&8 & 12 \\ 
$\pi$ Sco       & B1 V+...       &347&2 &  20&2 & 2 \\ 
$\delta$ Sco    & B0.2 IVe       &350&1 &  22&5 & 9 \\
\cutinhead{Other masked regions}
\multicolumn{2}{l}{Gum Nebula}				& 255&8 &  -9&1 & $\sim 40$ \\
\multicolumn{2}{l}{Orion-Eridanus bubble}	& 45&	& -25&	& $\sim 25$ \\
\enddata

\tablecomments{Ionizing stars of \ion{H}{2} regions which we removed from the WHAM Northern Sky Survey to leave a sample of only the diffuse WIM. This list includes only high-latitude ($|b| > 10 \arcdeg$) regions. Some sightlines are contained in multiple, overlapping \ion{H}{2} regions. The radius removed is the maximum extent of significantly enhanced \Ha\ and may be larger than the size of asymmetrical \ion{H}{2} regions. Coordinates and spectral types are from the SIMBAD Astronomical Database (\url{http://simbad.u-strasbg.fr/}); see also Table~1 of \citet{mrh06}.}
\tablenotetext{a}{See \citet{h01} for a discussion of the ionizing sources in this region.}
\label{tbl:hii_regions}

\end{deluxetable}

\begin{figure*}
\plottwo{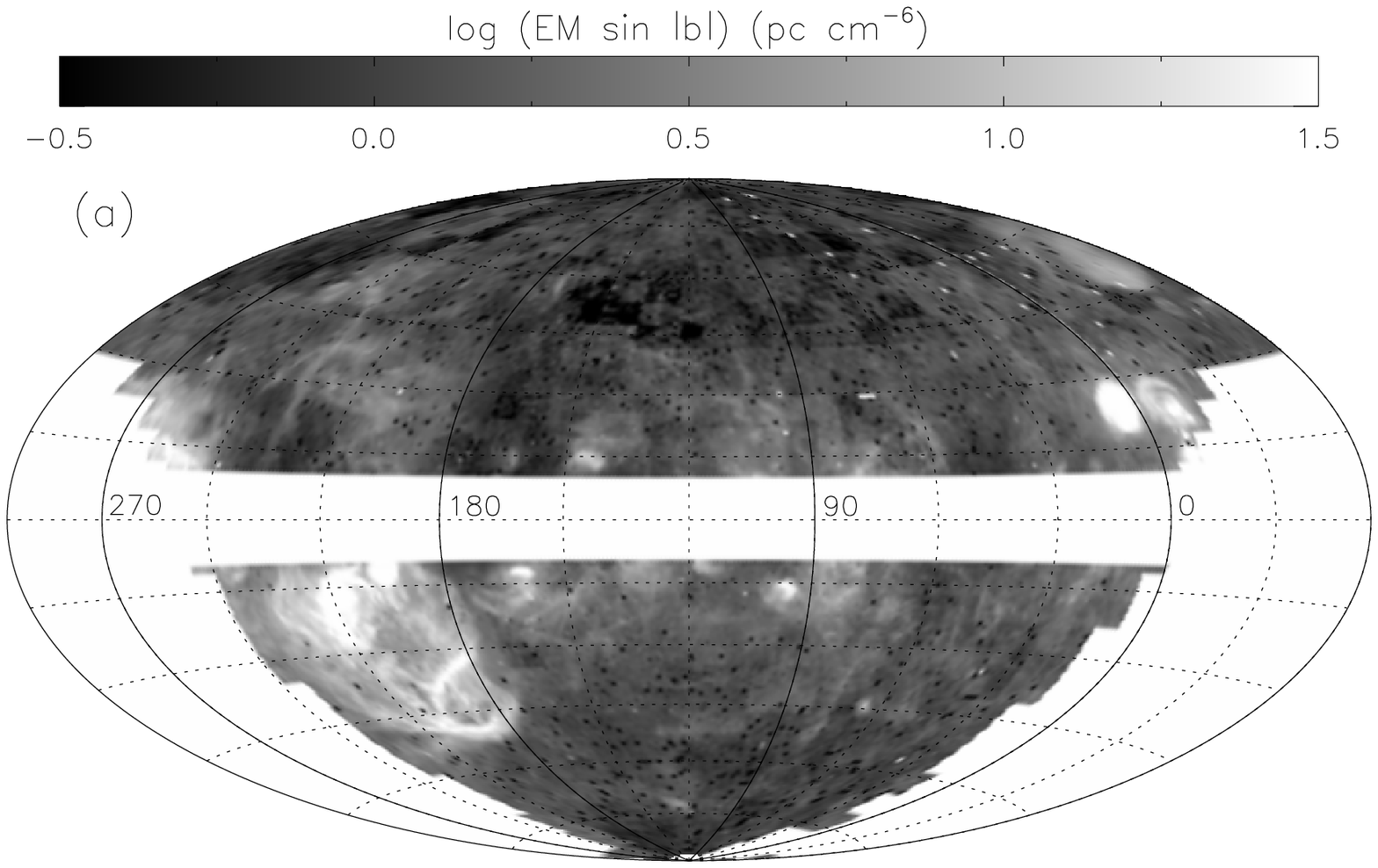}{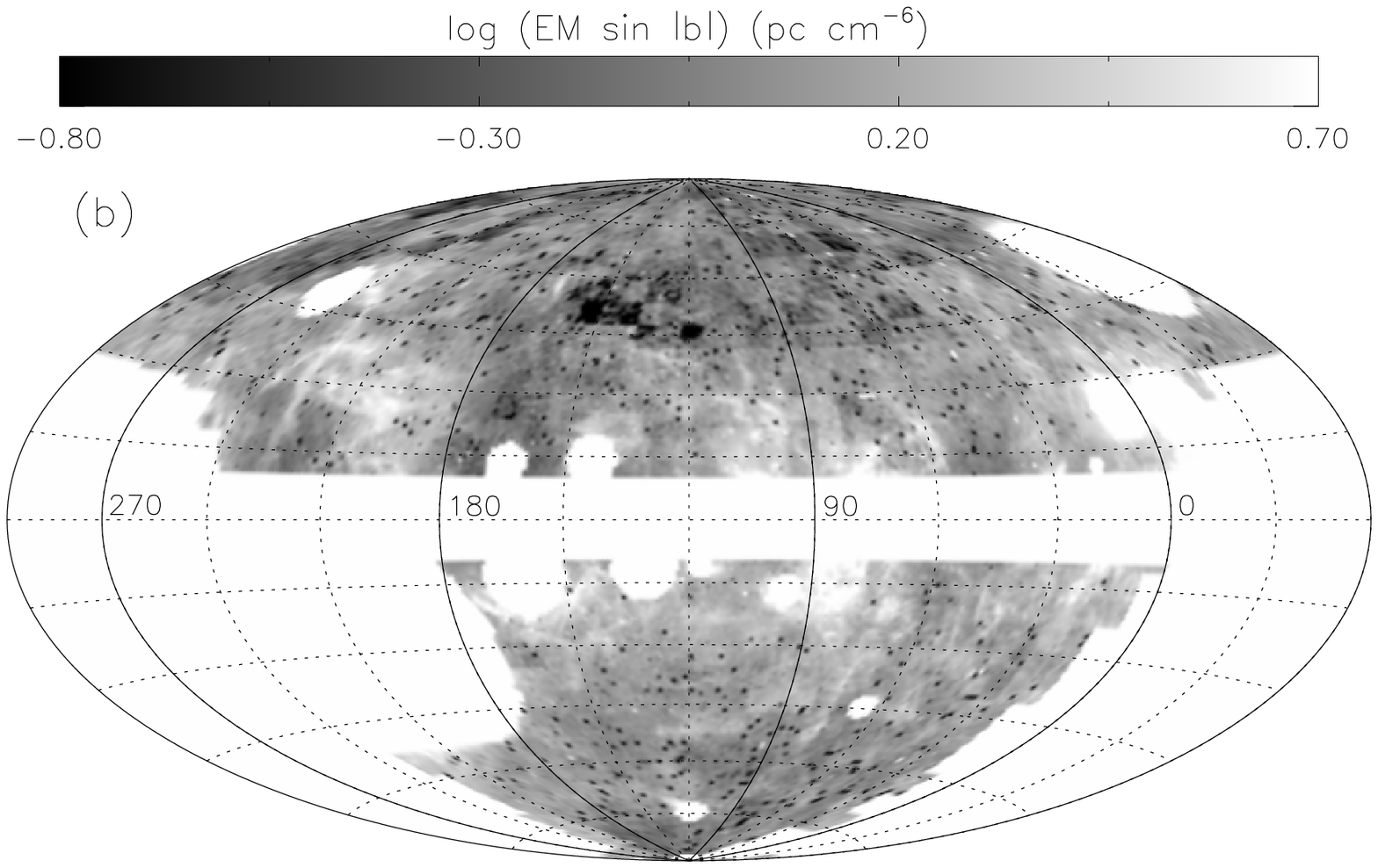}
\caption{Vertical component of \Ha\ emission measure \emsinb\ from the WHAM Northern Sky Survey for all $|b| > 10 \arcdeg$ sightlines (panel $a$) and with sightlines intersecting classical \ion{H}{2} regions removed (panel $b$). Maps are centered on $l = 120 \arcdeg$; dotted lines represent $15 \arcdeg$ increments in latitude and $30 \arcdeg$ increments in longitude. Single pixel dark spots are sightlines contaminated by bright stars which were removed in the processing of the survey.}
\label{fig:sky_map}
\end{figure*}

\begin{figure}
\plotone{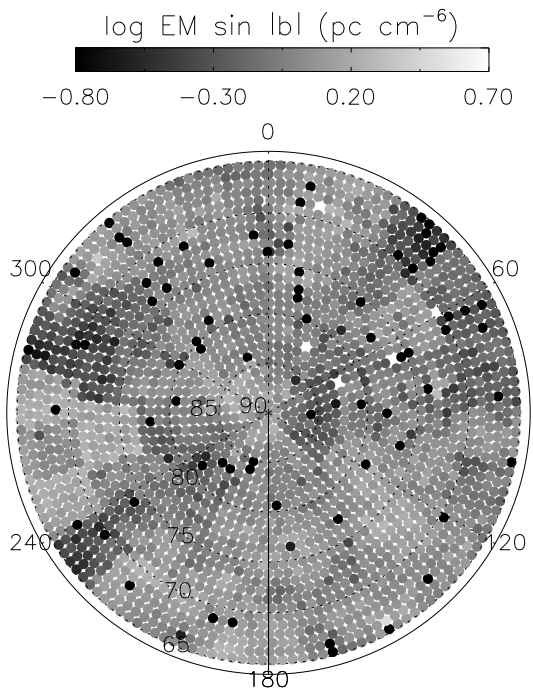}
\caption{Map of \emsinb\ from the WHAM survey for the northern polar region. Rectangular regions with discontinuities in intensity represent $7 \times 7 \arcdeg$ observational ``blocks'' which were observed together; at the low intensities observed at high latitudes \citep{hrt03}, block-to-block changes in the sky subtraction represent a significant fraction of the emission.}
\label{fig:polar_map}
\end{figure}

The Wisconsin H-Alpha Mapper (WHAM) Northern Sky Survey \citep{hrt03} is a spectroscopical map of \Ha\ emission in the northern sky within $\pm 100 \kms$ of the local standard of rest with an angular resolution of $1 \arcdeg$. The spectra were obtained using a dual-etalon Fabry-Perot spectrometer with a spectral resolution of $12 \kms$ and a sensitivity of $0.15$~R ($0.34 \pc \cmsix$). The spectral resolution of the WHAM survey allows the subtraction of numerous atmospheric emission lines, including the relatively bright geocoronal \Ha\ line, to establish a stable zero point for the interstellar \Ha\ intensities.

We calculated emission measures from the WHAM survey integrated over velocities $|v| < 100 \kms$ according to \citep{r91}
\begin{eqnarray} \label{eq:em}
\EM = \int n_e^2 \, ds  
	= 2.75 \left(\frac{T}{10^4 \K} \right)^{0.9} \left(\frac{I_{\Ha}}{1 \textrm{ R}} \right) \textrm{ pc cm}^{-6},
\end{eqnarray}
where $n_e$ is the local electron density, $I_{\Ha}$ is the \Ha\ intensity ($1 \textrm{ R} = 10^6 (4\pi)^{-1} \textrm{ photons cm}^{-2} \textrm{ s}^{-1} \textrm{ sr}^{-1}$), and $T$ is the temperature. To determine the temperature, we consider the ionization states in the WIM. Through the WIM, \sii$/$\nii\ line ratios are relatively constant, which implies a uniform ionization state, but \sii$/$\Ha\ and \nii$/$\Ha\ line ratios tend to be higher in lower \Ha\ intensity regions and at higher $|z|$ \citep{hrt99}. In combination, these observations imply a range of temperatures in the WIM from $6000 - 10000 \K$, with higher temperatures higher above the plane and in low density regions. We adopt a temperature of  $8000 \K$.

Within about $5 \arcdeg$ of the midplane, extinction by dust obscures many sightlines. Therefore, we restrict our data set to sightlines with Galactic latitude $|b| > 10 \arcdeg$, where extinction is negligible at \Ha\ \citep{hrt03, mr05}. Figures~\ref{fig:sky_map}$a$ and \ref{fig:polar_map} show a map of \emsinb, the emission measure corrected for path length through the WIM.

\subsection{Removed \ion{H}{2} region sightlines} \label{sec:hii_removal}

In order to sample only the WIM, we have removed from the WHAM survey sightlines that intersect \Ha\ emission due to identifiable, discrete structures with known sources of ionizing radiation. We identified 17 large, classical \ion{H}{2} regions with $|b| > 10\arcdeg$ in the WHAM survey, then removed sightlines within a circle centered on the ionizing star. We chose the radius of the circle by eye to encompass the \Ha\ enhancement of the \ion{H}{2} region. Because many regions are not circular, this crude mask removes some sightlines that may properly belong to the diffuse WIM. We list the masked \ion{H}{2} regions in Table~\ref{tbl:hii_regions}. We also removed two regions with significantly enhanced \Ha\ emission which are not classical \ion{H}{2} regions. The \object{Gum Nebula} is a large \ion{H}{2} region centered at $l = 258 \arcdeg$ and $b = -5 \arcdeg$ with a diameter of $36 \arcdeg$. Most of the Gum Nebula is below declination $\delta = -30 \arcdeg$, the southern limit of the WHAM survey. The \Ha\ emission from the \object{Orion-Eridanus super-bubble} ($180 \arcdeg \lesssim l \lesssim  270 \arcdeg$; $-50 \arcdeg \lesssim b \lesssim 0 \arcdeg$) is discussed in detail by \citet{ro79, hhr99, hhr00}; and \citet{mrh06}.

The remaining portion of the WHAM survey has $26,221$ sightlines which sample primarily the WIM---plasma with no readily identifiable, discrete source of ionizing radiation. A map of these diffuse WIM sightlines is shown in Figure~\ref{fig:sky_map}$b$. Considerable structure remains evident, including high-latitude filaments \citep[e.\ g.][]{hrt98}. \citet{mrh06} found that the line ratios \nii$/$\Ha\ (which is sensitive to the temperature of the gas) and \sii$/$\Ha\ in three discrete high-latitude filaments are different from line ratios observed in classical \ion{H}{2} regions and are indistinguishable from the ratios in the WIM. Therefore, we consider the high-latitude filaments as part of the WIM sample. We did not remove ``WHAM Point Sources'' \citep{rcm05}---$1$ pixel enhancements in the WHAM survey, in many cases asscoiated with planetary nebulae---because the 84 identified point sources contaminate only 91 sightlines and the excess emission measures are much smaller than those of the classical \ion{H}{2} regions.

\section{Distribution of emission measure} \label{sec:dist}

\begin{figure}
\plotone{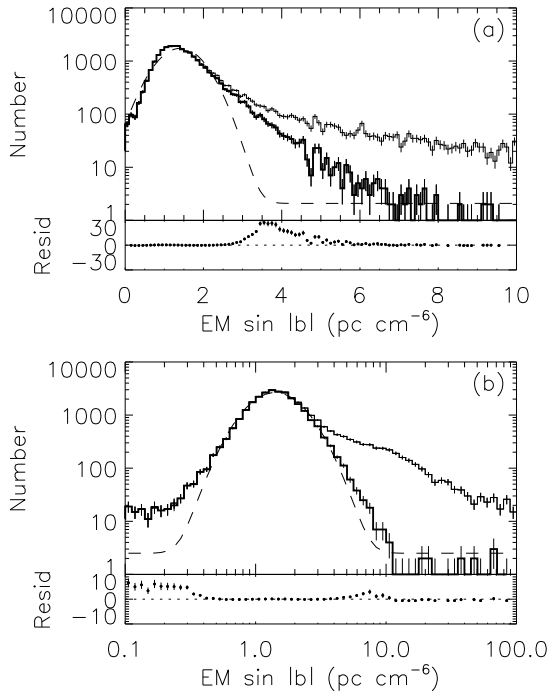}
\caption{Histograms of \emsinb\ for all $|b| > 10 \arcdeg$ sightlines in the WHAM survey (thin, solid lines) and for diffuse WIM sightlines (thick, solid lines). Gaussian and lognormal fits to the diffuse WIM distribution are shown with dashed lines. Neither sample is well fit by a normal distribution (panel $a$), but the WIM is well matched by a lognormal distribution (panel $b$). Residuals comparing each fit to the diffuse WIM distribution are shown in the lower panel. They are defined as $(N_\mathrm{WIM} - N_\mathrm{fit})/N_\mathrm{fit}$, where $N_\mathrm{WIM}$ is the value in each histogram bin, are shown below for the WIM distribution.}
\label{fig:emsinb_hist}
\end{figure}

\begin{deluxetable}{c r@{.}l@{$\pm$}r@{.}l r@{.}l@{$\pm$}r@{.}l r r@{.}l}

\tabletypesize{\footnotesize}
\tablecolumns{12}
\tablewidth{0pt}
\tablecaption{Lognormal Fits of $\EM \sin |b|$}

\tablehead{
    \colhead{range} & \mcf{$10^{\langle \log \emsinb \rangle}$} & \mcf{$\sigma_{\log \emsinb}$} & \colhead{$N$} & \mc{$\tilde{\chi}^2$} \\
    \colhead{} & \mcf{$(\pc \cmsix)$} & \mcf{(dex)} & \colhead{} & \mc{}
}

\startdata

Full WIM & 1&406 & 0&004 & 0&190 & 0&001 & 26221 &  3&9 \\
\hline
$ 10 < |b| <  30$ & 1&798 & 0&009 & 0&192 & 0&002 &  8747 &  1&6 \\
$ 10 < b <  30$ & 1&647 & 0&011 & 0&200 & 0&002 &  5439 &  1&3 \\
$-30 < b < -10$ & 1&979 & 0&012 & 0&135 & 0&002 &  3308 &  4&0 \\
\hline
$ 30 < |b| <  60$ & 1&356 & 0&005 & 0&156 & 0&001 & 12296 &  3&9 \\
$ 30 < b <  60$ & 1&316 & 0&006 & 0&177 & 0&002 &  8216 &  2&9 \\
$-60 < b < -30$ & 1&409 & 0&006 & 0&115 & 0&001 &  4080 &  1&8 \\
\hline
$ 60 < |b| <  90$ & 1&068 & 0&005 & 0&139 & 0&002 &  5178 &  3&0 \\
$ 60 < b <  90$ & 0&991 & 0&006 & 0&132 & 0&002 &  3246 &  3&8 \\
$-90 < b < -60$ & 1&198 & 0&009 & 0&127 & 0&003 &  1932 &  0&7 \\

\enddata

\tablecomments{Gaussian fits to the distribution of $\log \EM \sin |b|$ for subsets of all WIM sightlines -- the WHAM survey at $|b| > 10 \arcdeg$ with classical \ion{H}{2} regions removed. There are $N$ pointings in each region; $\tilde{\chi}^2$ is the reduced $\chi^2$ goodness-of-fit parameter. See Figs.~\ref{fig:emsinb_hist} and \ref{fig:emhist_lat}.}
\label{tbl:emhist_fit}

\end{deluxetable}

Histograms of  \emsinb\ are shown in Figure~\ref{fig:emsinb_hist}$a$. The histogram for the entire high-latitude ($|b|>10 \arcdeg$) WHAM survey shows a peak near $\emsinb = 1.4 \pc \cmsix$. When this histogram is plotted with a linear \emsinb\ axis, the distribution shows an extended tail to large values of \emsinb.  Restricting the sample to the WIM sightlines reduces this high \emsinb\ tail, but does not eliminate it. The figure also shows that the \emsinb\ histogram is clearly not well fit by a normal (Gaussian) distribution; at high \emsinb, the data can be as much as $27$ times greater than the best-fit Gaussian.

However, when the \emsinb\ distribution is plotted on a logarithmic scale for the horizontal \emsinb\ axis (Figure~\ref{fig:emsinb_hist}$b$), we find that the distribution is {\em extremely} well fit by a lognormal distribution of \emsinb\ or, equivalently, a normal distribution of $\log \emsinb$. This is particularly true for the pure WIM sample, where the greatest deviation between the data and the best fit lognormal distribution at high \emsinb\ is a factor of $3.0$. In fact, the fit is so good that a comparison of a lognormal distribution to the observed \emsinb\ distribution is a useful indicator of the quality of the chosen WIM sample. With an incomplete removal of \ion{H}{2} regions, the fit to a lognormal distribution is poor at high emission measures. Parameters of single-component fits of the distribution of $\log \emsinb$ for WIM sightlines are listed in Table~\ref{tbl:emhist_fit}. The reduced $\chi^2$ goodness-of-fit parameter \citep[e.~g.][]{t82},
\begin{equation}
\tilde{\chi}^2 = \sum_{\mathrm{bins} \, i} \left( \frac{N_i - f(\emsinb_i)}{\sigma_{Ni}} \right)^2 / d,
\end{equation}
is also listed, where $N_i$ is the number of pointings in bin $i$, $\sigma_{Ni} = \sqrt{N_i}$ is the Poisson error, $f(\emsinb_i)$ is the modeled count in each bin, and $d$ is the number of degrees of freedom in the fit. We fit Gaussians with a zero offset parameter, so each fit has four constraints.

\subsection{Variations with latitude} \label{sec:dist_lat}

\begin{figure}
\plotone{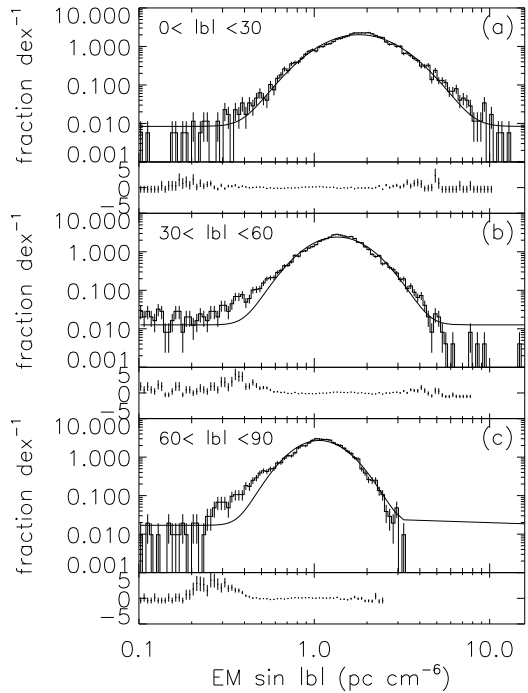}
\caption{Histograms of $\log \emsinb$ for WIM sightlines in a range of Galactic latitudes with Poisson errors. Lognormal fits and residuals (as defined in Fig.~\ref{fig:emsinb_hist}) are shown; fit parameters are shown in Table~\ref{tbl:emhist_fit}.}
\label{fig:emhist_lat}
\end{figure}

We found no evidence for any significant change of the  \emsinb\ distribution as a function of Galactic longitude, but there is a dependence on Galactic latitude. This trend is visible in the all-sky map (Fig.~\ref{fig:sky_map}) which shows that  \emsinb\ appears to be lower at high latitudes than at low latitudes. The distributions of \emsinb\ in three latitude bins are plotted in Figure~\ref{fig:emhist_lat}; lognormal fit parameters are given in Table~\ref{tbl:emhist_fit}. The center of the distribution of \emsinb\ is $\sim 0.2$~dex lower at high ($|b| > 60 \arcdeg$) than at low ($|b| < 30 \arcdeg$) latitude. The distribution is also $\sim 0.05$~dex wider at low latitude, although this trend is not present at negative latitudes, which are less completely sampled by the WHAM Northern Sky Survey than positive latitudes.

In a plane-parallel medium, \emsinb\ would be constant with latitude. The observed variation with latitude may indicate that the Sun is in a relatively low-density region of the Galaxy because lower-latitude sightlines sample longer pathlengths within the plane than high-latitude sightlines \citep{r97}. For example, a $b = 10 \arcdeg$ sightline samples a cylinder of radius $5.7$~kpc before $d \sin |b|$ is greater than the 1~kpc scale height of the WIM, whereas a $b = 60 \arcdeg$ sightline samples only within a $0.6$~kpc cylinder around the Sun. This low density region of ionized gas in the solar environs is probably not primarily due to the Local Bubble, which is much smaller ($\sim 100 \pc$ in radius) than the scale height of the WIM and thus only accounts for a small fraction of the line of sight. Additionally, some lower latitude sightlines include spiral arms, which have higher densities than the more local gas sampled by high-latitude sightlines \citep{hrt03}.

\section{Turbulence in the warm ionized medium} \label{sec:turbulence}

What is the physical mechanism that produces the observed lognormal distribution of \emsinb? If one starts with an average density for the WIM and repeatedly adds or subtracts a small increment to each volume element, the central limit theorem shows that the resulting probability distribution function (PDF) of density will be normal (Gaussian). Similarly, a series of small compressions and rarefactions, which is  expected for a compressible fluid, will produce a lognormal distribution because multiplication is additive in logarithmic units \citep{v94,pv98,vp99}. Thus it is quite natural to expect a lognormal density distribution in the WIM. Models of turbulence, for example, produce a lognormal distribution of density \citep{pjn97,np99,ogs99,osg01}. In the presence of magnetic pressure, the lognormal distribution of density is not exact. However, at large \Ma, the magnetic pressure decorrelates from density, leaving a lognormal density PDF \citep{pv03}. The effect of magnetic fields on the PDF is also small when shocks in the medium are randomly oriented with respect to the field \citep{blc05}.

If the path length of a column is long enough for elements within the column to be uncorrelated, the central limit theorem implies that the column density distribution should approach a Gaussian. However, if the turbulence is correlated on a scale longer than the path length, the column density is expected to have the same distribution function as the density \citep{vg01}, as is seen numerically \citep{osg01}. Because $\log(n^2) = 2 \log n$, if the column density ($\int n \, ds$) is lognormally distribued, we expect the emission measure ($\int n^2 \, ds$) to be lognormally distributed as well. \citet{vg01} define the ratio of the path length to the decorrelation length as the parameter $\eta$, with full convergence to a Gaussian column density distribution expected for $\eta \gg 1$ and partial convergence for $\eta \gtrsim 1$. The decorrelation length is not known {\em a priori}, although \citet{vg01} suggest that it may be related to the outer scale.

The outer scale of the turbulence in the WIM in the Milky Way is not known, with estimates ranging from $4 \pc$ \citep{ms96} to $30-100 \pc$ \citep[e.\ g.][]{ars95, db07}. \citet{cls06} apply the velocity coordinate spectrum \citep{lp06} technique to \ion{H}{1} data and find an injection scale of 105~pc. Studies of \ion{H}{1} in dwarf irregular galaxies such as the \object{Large} and \object{Small Magellanic Cloud}s and \object{Holmberg II} have found outer scales of several kiloparsecs, much larger than most claims within the Milky Way \citep{ssd99,eks01,db05}. \citet{hgb06,hbg08} suggest that the outer scale in the magneto-ionized medium in spiral arms is quite small (a few parsecs) with the turbulence related to \ion{H}{2} regions, while the global magneto-ionized ISM has a much larger outer scale ($\sim 100 \pc$) with turbulence driven by supernovae. At a distance of $1 \kpc$, the $1 \arcdeg$ beam of WHAM corresponds to a transverse length of $17 \pc$, so our data are not sensitive to parsec-scale fluctuations.
\notetoeditor{``Large'' in the object tag refers to the Large Magellanic Cloud}

\section{MHD models} \label{sec:model}

We have chosen to compare our observational results to models of MHD isothermal turbulence. The WIM is a nearly fully ionized plasma and is observed to be magnetized. The isothermal assumption is justified because the temperatures of the WIM are observed to lie in the range $6000 - 10000 \K$ \citep{hrt99}, so the sound speed varies by $< 30 \%$ within the medium. \citep[hereafter KLB]{klb07} developed simulations of 3D compressible, isothermal MHD turbulence over many dynamical times for an extended range of sonic and Alfv\'en Mach numbers. Although these models were not developed with the WIM in mind, they provide a useful context for interpreting our observations.

For a plasma with thermal pressure $p$, mass density $\rho$,  and magnetic field strength $B$, each simulation is principally characterized by two numbers: the sonic Mach number, $\Ms = v / c_s$, and the Alfv\'en Mach number, $\Ma = v / c_A$, where $v$ is the particle speed in each cell. The sound speed and Alfv\'en speed are given by  $c_s = \sqrt{\partial p / \partial \rho}$ and $c_A = B / \sqrt{4 \pi \rho}$, respectively. The models are described in detail by KLB and references therein. The turbulence was driven using a solenoidal forcing function in Fourier space; this minimizes the impact of the chosen forcing function on the resulting density structures.

KLB analyzed the statistics of 3D density and 2D column density, including PDFs, spectra, structure functions, and intermittency of density and the logarithm of density, in order to establish the relation between the statistics of the observables (such as column densities) and the underlying 3D statistics of density.  They found that the amplitude of density fluctuations strongly depends on the sonic Mach number in both weakly and strongly magnetized turbulent plasmas. Moreover, PDFs of density fluctuations have a lognormal distribution with a width growing with \Ms.
The KLB models have $\eta$ ranging from 5 (for subsonic models) to 13 (supersonic), meaning only partial convergence to a Gaussian column density PDF is expected; the column density distributions of the models are very nearly lognormal, like the underlying density distributions.

We match the simulations presented by KLB to the physical conditions of the WIM and compare the resulting distributions of simulated emission measure and column density to WHAM and pulsar data. We use $256 \times 256 \times 256$ element simulations in both super- and sub-Alfv\'enic and super- and sub-sonic regimes. Most of the models analyzed here were presented in KLB, but we added two mildly supersonic ($\Ms = 1.4 - 1.7$) models for this work. Most models were run for at least $10$ dynamical times and all reached approximate steady-states in all of the statistical parameters.

\subsection{Setting physical scales} \label{sec:normalization}

\begin{\deluxetablestar}{l r@{.}l r@{.}l r@{.}l r@{.}l@{$\pm$}r@{.}l r@{.}l@{$\pm$}r@{.}l r@{.}l@{$\pm$}r@{.}l r@{$\pm$}r }

\tabletypesize{\footnotesize}
\tablecolumns{21}
\tablewidth{0pt}
\tablecaption{Globular cluster pulsar and emission measure data}

\tablehead{
\colhead{Globular} & \mc{$l$} & \mc{$b$} & \mc{$z$} & \mcf{$\textrm{DM} \sin |b|$} & \mcf{$\textrm{EM} \sin |b|$} & 
	\mcf{$n_c$} & \mc{$L_c \sin |b|$} \\
\colhead{Cluster} & \mc{($\arcdeg$)} & \mc{($\arcdeg$)} & \mc{(kpc)} & \mcf{($\textrm{pc cm}^{-3}$)} & \mcf{$(\textrm{pc cm}^{-6})$} &
	\mcf{($\textrm{cm}^{-3}$)} & \mc{(pc)}
}

\startdata
 M3 &  42&3 &  78&7 &  10&2 &   25&89 &   0&05 &  0&95 & 0&03 & 0&037 & 0&001 &   703 &    22 \\
 M5 &   4&0 &  46&8 &   5&7 &   21&66 &   0&03 &  1&31 & 0&02 & 0&060 & 0&001 &   359 &     5 \\
M13 &  59&1 &  40&9 &   4&8 &   19&87 &   0&02 &  1&12 & 0&06 & 0&056 & 0&003 &   352 &    19 \\
M15 &  65&1 & $-27$&3 & $-4$&1 & 30&812 & 0&014 & 3&54 & 0&04 & 0&115 & 0&001 &   268 &     3 \\
M30 &  27&2 & $-46$&8 & $-6$&5 & 18&283 & 0&003 & 1&94 & 0&07 & 0&106 & 0&004 &   172 &     6 \\
M53 & 333&1 &  79&8 &  16&9 &    23&6 &    1&5 &  1&14 & 0&07 & 0&048 & 0&004 &   488 &    53 
\enddata
\tablecomments{Properties of lines of sight towards $|z| > 3$ kpc globular clusters with known pulsars. EM data from the WHAM survey; sightlines towards M3 and M5 are contaminated by bright stars (see text). DM and distance data from the ATNF pulsar catalog \citep{mht05}.}
\label{tbl:gc_data}
\end{\deluxetablestar}

Since the simulations were performed in dimensionless coordinates, it is necessary to use three parameters to specify the physical dimensions of the simulation. We describe this procedure in detail in Appendix~\ref{app}. Once the physical parameters of each simulation are uniquely specified, we can determine how well each simulation matches the observed emission measure PDF as well as other observations. As described below, we have chosen to fix the following three parameters: (1) total vertical dispersion measure through the box, $\DM \sin |b|=23 \pc \cucm$, (2) length of the box, $\hsimbox =200 - 1000 \pc$, and (3) temperature of the gas, $T=8000 \K$ (see \S~\ref{sec:data}). 

Dispersion measures of pulsars more than $\sim 3$ scale heights above the midplane sample gas from the entire WIM, so we use the value of $\DM \sin |b|$ for these pulsars to set the column density through the simulation cube. In practice, the only pulsars high above the midplane with reliable distance measurements are those in globular clusters. Six globular clusters with $|z| > 3 \kpc$ contain known pulsars \citep{mht05}. A recent search for pulsars in all globular clusters visible from Arecibo failed to detect pulsars in additional clusters at $|z| > 3 \kpc$ \citep{hrs07}. The mean $\DM \sin |b|$ for the $|z| > 3 \kpc$ globular clusters is $23 \pc \cucm$ (Table~\ref{tbl:gc_data}). 

Dispersion measures of pulsars with \DM-independent distance measurements establish the scale height of the WIM as $\hwim = 1 \kpc$ \citep{r89a, r97, gbc01}; \Ha\ intensities towards the Perseus arm yield the same result \citep{hrt99}. A recent analysis of high-latitude ($|b| > 40 \arcdeg$) pulsars with distance measurements from either parallax or association with a globular cluster found a considerably higher scale height for the WIM of $1.8 \kpc$ \citep{gmc08}. Because this recent study samples primarily gas within a $\approx 2 \kpc$ cylinder around the Sun, it may imply a different WIM scale height locally than in spiral arms. Comparisons of \EM\ and \DM\ have indicated that the WIM occupies $\sim 20-40 \%$ of the volume within the $2 \kpc$ thick layer about the midplane \citep[e.\ g.][but see \S~\ref{sec:f}]{r91}. Combinations of emission measures and dispersion measures and dispersion measures and the scale height both constrain the number density of the gas, so the problem is overconstrained. Therefore, we use simulation box sizes ranging from $200 - 1000 \pc$ and test each scale by comparing simulated emission measures to the WHAM data. Given that the actual occupied volume of the WIM must be no greater than the observed scale height, we do not test box sizes $>1.0 \kpc$ even though the scale height could be larger. The solenoidal forcing function driving the turbulence has a wavelength one-quarter the length of the simulation box, which corresponds to $50-250 \pc$.

\subsection{Determining the best-fitting model} \label{sec:bestfit}

\begin{figure*}
\plotone{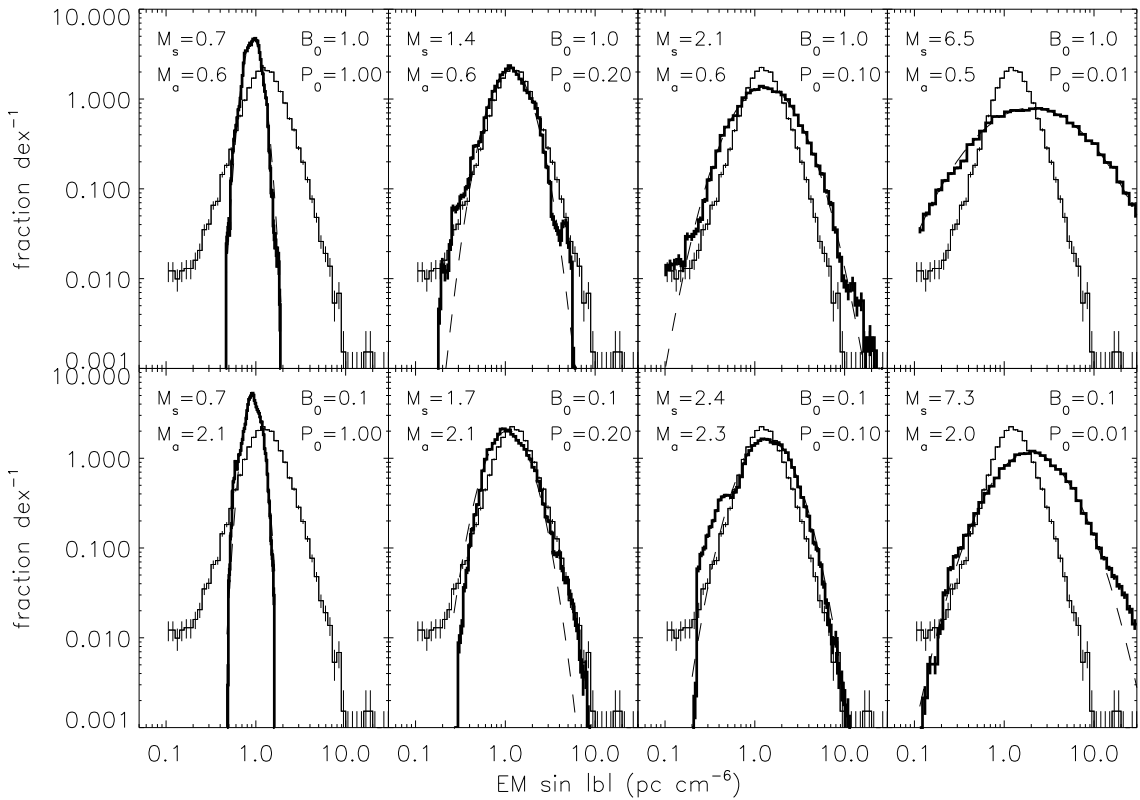}
\caption{Histograms of \Ha\ emission measure \emsinb\ from WIM sightlines in the WHAM northern sky survey ({\em thin lines}) and simulated \EM\ from MHD simulations with a box size of $500 \pc$ ({\em thick lines}). Lognormal fits of the simulated \EM\ distribution are also shown ({\em dashed lines}). Initial external magnetic field strengths and initial pressures computed from each model are specified in code units, and sonic and Alfv\'enic Mach numbers are also listed.}
\label{fig:kowal_em}
\end{figure*}

\begin{figure}
\plotone{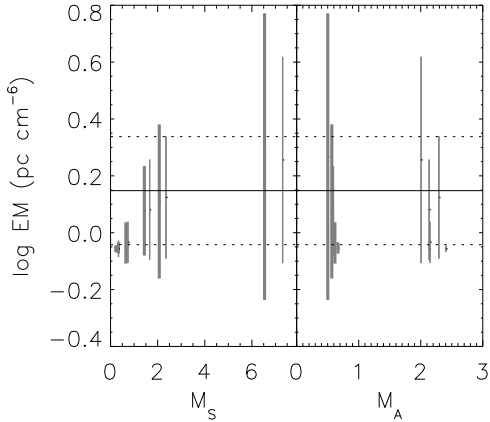}
\caption{Comparsion of observed and modeled \emsinb\ ranges as a function of both sonic and Alfv\'en Mach number. We use a $500 \pc$ box size in each case.  Vertical gray bars depict the $\pm 1 \sigma$ range of Gaussian fits to the $\log(\emsinb)$ distributions of each of the twelve models considered. The thinner bars correspond to the weakly magnetized (high \Ma) cases; the thicker bars correspond to the strongly magnetized (low \Ma) cases.  The left panel shows both the mean and spread increasing as \Ms\ increases. The solid and dashed horizontal lines show the observed mean ($\langle \log \left( \emsinb \, (\mathrm{pc} \cmsix)^{-1} \right) \rangle = 0.146$) and standard deviation ($0.190$) of the distribution.  The mean of the simulated distributions is set by our physical scaling of the simulation (\S~\ref{sec:normalization}), but the width is independent of choice of scaling parameters. Error bars are omitted for clarity.}
\label{fig:ms_ma_em}
\end{figure}

\begin{\deluxetablestar}{r@{.}l r@{.}l r@{.}l r@{.}l r@{.}l r r@{.}l r@{.}l r@{.}l r@{.}l r@{.}l r@{.}l r@{.}l r r}


\tabletypesize{\footnotesize}
\tablecolumns{27}
\tablewidth{0pt}
\tablecaption{Model fit parameters}

\tablehead{
	\mc{\Bext} & \mc{$\langle B \rangle$} & \mc{\Pini} & \mc{\Ms} & \mc{\Ma} & \colhead{$\hsimbox$} & \mc{$10^{\langle \log \EM \rangle}$} & \mc{$\sigma_{\log \EM}$} & \mc{$\tilde{\chi}^2$} & \mc{$10^{\langle \log n \rangle}$} & \mc{$\sigma_{\log n}$} & \mc{$\langle n_c \rangle$} & \mc{$\sigma_{n_c}$} & \colhead{$\langle L_c \rangle$} & \colhead{$\sigma_{L_c}$} \\
	\mc{} & \mc{($\mu \mathrm{G}$)} & \mc{} & \mc{} & \mc{} & \colhead{(pc)} & \mc{(pc \cmsix)} & \mc{(dex)} & \mc{} & \mc{(\cucm)} & \mc{(dex)} & \mc{(\cucm)} & \mc{(\cucm)} & \colhead{(pc)} & \colhead{(pc)}	\\
	\mc{(1)} & \mc{(2)} & \mc{(3)} & \mc{(4)} & \mc{(5)} & \colhead{(6)} & \mc{(7)} & \mc{(8)} & \mc{(9)} & \mc{(10)} & \mc{(11)} & \mc{(12)} & \mc{(13)} & \colhead{(14)} & \colhead{(15)} \\ \hline
	\multicolumn{11}{l}{Data (Tables \ref{tbl:emhist_fit} \& \ref{tbl:gc_data}):}    &  1&406 & 0&190 & \mc{\nodata} & \mcf{\nodata} & 0&07 & \mc{\nodata} & 300 & \nodata
}

\startdata
         1&0 &   1&79 &         1&00 &         0&65 &         0&62 &    200 &   2&315 &        0&085 &  33&7 &  0&104 &        0&113 &  0&111 &  0&011&    188 &     6 \\
\mc{\nodata} &   1&46 & \mc{\nodata} & \mc{\nodata} & \mc{\nodata} &    300 &   1&544 & \mc{\nodata} &  33&1 &  0&069 & \mc{\nodata} &  0&074 &  0&007&    283 &     9 \\
\mc{\nodata} &   1&27 & \mc{\nodata} & \mc{\nodata} & \mc{\nodata} &    400 &   1&158 & \mc{\nodata} &  32&5 &  0&052 & \mc{\nodata} &  0&055 &  0&005&    377 &    12 \\
\mc{\nodata} &   1&13 & \mc{\nodata} & \mc{\nodata} & \mc{\nodata} &    500 &   0&926 & \mc{\nodata} &  32&3 &  0&042 & \mc{\nodata} &  0&044 &  0&004&    472 &    15 \\
\mc{\nodata} &   1&03 & \mc{\nodata} & \mc{\nodata} & \mc{\nodata} &    600 &   0&772 & \mc{\nodata} &  32&9 &  0&035 & \mc{\nodata} &  0&037 &  0&004&    566 &    18 \\
\mc{\nodata} &   0&96 & \mc{\nodata} & \mc{\nodata} & \mc{\nodata} &    700 &   0&662 & \mc{\nodata} &  32&9 &  0&030 & \mc{\nodata} &  0&032 &  0&003&    661 &    22 \\
\mc{\nodata} &   0&89 & \mc{\nodata} & \mc{\nodata} & \mc{\nodata} &    800 &   0&579 & \mc{\nodata} &  32&8 &  0&026 & \mc{\nodata} &  0&028 &  0&003&    755 &    25 \\
\mc{\nodata} &   0&80 & \mc{\nodata} & \mc{\nodata} & \mc{\nodata} &   1000 &   0&463 & \mc{\nodata} &  33&0 &  0&021 & \mc{\nodata} &  0&022 &  0&002&    944 &    31 \\
\hline
         0&1 &   0&58 &         1&00 &         0&74 &         2&15 &    200 &   2&303 &        0&082 &  42&0 &  0&104 &        0&101 &  0&110 &  0&011&    190 &     4 \\
\mc{\nodata} &   0&48 & \mc{\nodata} & \mc{\nodata} & \mc{\nodata} &    300 &   1&535 & \mc{\nodata} &  41&9 &  0&069 & \mc{\nodata} &  0&073 &  0&007&    285 &     6 \\
\mc{\nodata} &   0&41 & \mc{\nodata} & \mc{\nodata} & \mc{\nodata} &    400 &   1&151 & \mc{\nodata} &  44&0 &  0&052 & \mc{\nodata} &  0&055 &  0&006&    381 &     8 \\
\mc{\nodata} &   0&37 & \mc{\nodata} & \mc{\nodata} & \mc{\nodata} &    500 &   0&921 & \mc{\nodata} &  41&2 &  0&042 & \mc{\nodata} &  0&044 &  0&004&    476 &    11 \\
\mc{\nodata} &   0&34 & \mc{\nodata} & \mc{\nodata} & \mc{\nodata} &    600 &   0&768 & \mc{\nodata} &  50&6 &  0&035 & \mc{\nodata} &  0&037 &  0&004&    571 &    13 \\
\mc{\nodata} &   0&31 & \mc{\nodata} & \mc{\nodata} & \mc{\nodata} &    700 &   0&658 & \mc{\nodata} &  42&1 &  0&030 & \mc{\nodata} &  0&031 &  0&003&    666 &    15 \\
\mc{\nodata} &   0&29 & \mc{\nodata} & \mc{\nodata} & \mc{\nodata} &    800 &   0&576 & \mc{\nodata} &  68&6 &  0&026 & \mc{\nodata} &  0&027 &  0&003&    762 &    17 \\
\mc{\nodata} &   0&26 & \mc{\nodata} & \mc{\nodata} & \mc{\nodata} &   1000 &   0&461 & \mc{\nodata} &  43&2 &  0&021 & \mc{\nodata} &  0&022 &  0&002&    952 &    22 \\
\hline
         1&0 &   3&83 &         0&20 &         1&43 &         0&58 &    200 &   2&853 &        0&184 &  26&6 &  0&086 &        0&301 &  0&143 &  0&037&    148 &    18 \\
\mc{\nodata} &   3&13 & \mc{\nodata} & \mc{\nodata} & \mc{\nodata} &    300 &   1&902 & \mc{\nodata} &  26&4 &  0&058 & \mc{\nodata} &  0&095 &  0&025&    223 &    27 \\
\mc{\nodata} &   2&71 & \mc{\nodata} & \mc{\nodata} & \mc{\nodata} &    400 &   1&427 & \mc{\nodata} &  26&4 &  0&043 & \mc{\nodata} &  0&071 &  0&019&    297 &    36 \\
\mc{\nodata} &   2&42 & \mc{\nodata} & \mc{\nodata} & \mc{\nodata} &    500 &   1&141 & \mc{\nodata} &  26&5 &  0&035 & \mc{\nodata} &  0&057 &  0&015&    372 &    45 \\
\mc{\nodata} &   2&21 & \mc{\nodata} & \mc{\nodata} & \mc{\nodata} &    600 &   0&951 & \mc{\nodata} &  26&5 &  0&029 & \mc{\nodata} &  0&048 &  0&012&    446 &    54 \\
\mc{\nodata} &   2&05 & \mc{\nodata} & \mc{\nodata} & \mc{\nodata} &    700 &   0&815 & \mc{\nodata} &  26&4 &  0&025 & \mc{\nodata} &  0&041 &  0&011&    521 &    63 \\
\mc{\nodata} &   1&92 & \mc{\nodata} & \mc{\nodata} & \mc{\nodata} &    800 &   0&713 & \mc{\nodata} &  26&4 &  0&022 & \mc{\nodata} &  0&036 &  0&009&    595 &    72 \\
\mc{\nodata} &   1&71 & \mc{\nodata} & \mc{\nodata} & \mc{\nodata} &   1000 &   0&571 & \mc{\nodata} &  26&5 &  0&017 & \mc{\nodata} &  0&029 &  0&007&    744 &    90 \\
\hline
         0&1 &   1&23 &         0&20 &         1&66 &         2&13 &    200 &   2&666 &        0&196 &  65&0 &  0&090 &        0&263 &  0&142 &  0&038&    149 &    15 \\
\mc{\nodata} &   1&00 & \mc{\nodata} & \mc{\nodata} & \mc{\nodata} &    300 &   1&777 & \mc{\nodata} &  65&0 &  0&060 & \mc{\nodata} &  0&094 &  0&025&    224 &    22 \\
\mc{\nodata} &   0&87 & \mc{\nodata} & \mc{\nodata} & \mc{\nodata} &    400 &   1&333 & \mc{\nodata} &  65&7 &  0&045 & \mc{\nodata} &  0&071 &  0&019&    299 &    30 \\
\mc{\nodata} &   0&78 & \mc{\nodata} & \mc{\nodata} & \mc{\nodata} &    500 &   1&067 & \mc{\nodata} & 121&7 &  0&036 & \mc{\nodata} &  0&057 &  0&015&    374 &    37 \\
\mc{\nodata} &   0&71 & \mc{\nodata} & \mc{\nodata} & \mc{\nodata} &    600 &   0&888 & \mc{\nodata} &  65&7 &  0&030 & \mc{\nodata} &  0&047 &  0&013&    449 &    45 \\
\mc{\nodata} &   0&66 & \mc{\nodata} & \mc{\nodata} & \mc{\nodata} &    700 &   0&762 & \mc{\nodata} & 120&9 &  0&026 & \mc{\nodata} &  0&040 &  0&011&    524 &    53 \\
\mc{\nodata} &   0&61 & \mc{\nodata} & \mc{\nodata} & \mc{\nodata} &    800 &   0&667 & \mc{\nodata} &  57&8 &  0&022 & \mc{\nodata} &  0&035 &  0&010&    598 &    60 \\
\mc{\nodata} &   0&55 & \mc{\nodata} & \mc{\nodata} & \mc{\nodata} &   1000 &   0&533 & \mc{\nodata} &  73&6 &  0&018 & \mc{\nodata} &  0&028 &  0&008&    748 &    75 \\
\hline
         1&0 &   5&31 &         0&10 &         2&07 &         0&57 &    200 &   3&239 &        0&291 &   8&7 &  0&076 &        0&398 &  0&177 &  0&077&    124 &    26 \\
\mc{\nodata} &   4&33 & \mc{\nodata} & \mc{\nodata} & \mc{\nodata} &    300 &   2&160 & \mc{\nodata} &   8&9 &  0&051 & \mc{\nodata} &  0&118 &  0&051&    186 &    39 \\
\mc{\nodata} &   3&75 & \mc{\nodata} & \mc{\nodata} & \mc{\nodata} &    400 &   1&619 & \mc{\nodata} &   8&7 &  0&038 & \mc{\nodata} &  0&088 &  0&038&    249 &    52 \\
\mc{\nodata} &   3&36 & \mc{\nodata} & \mc{\nodata} & \mc{\nodata} &    500 &   1&295 & \mc{\nodata} &   8&7 &  0&030 & \mc{\nodata} &  0&071 &  0&031&    311 &    65 \\
\mc{\nodata} &   3&06 & \mc{\nodata} & \mc{\nodata} & \mc{\nodata} &    600 &   1&080 & \mc{\nodata} &   8&9 &  0&025 & \mc{\nodata} &  0&059 &  0&026&    373 &    79 \\
\mc{\nodata} &   2&84 & \mc{\nodata} & \mc{\nodata} & \mc{\nodata} &    700 &   0&925 & \mc{\nodata} &   8&7 &  0&022 & \mc{\nodata} &  0&050 &  0&022&    435 &    92 \\
\mc{\nodata} &   2&65 & \mc{\nodata} & \mc{\nodata} & \mc{\nodata} &    800 &   0&810 & \mc{\nodata} &   8&9 &  0&019 & \mc{\nodata} &  0&044 &  0&019&    498 &   105 \\
\mc{\nodata} &   2&37 & \mc{\nodata} & \mc{\nodata} & \mc{\nodata} &   1000 &   0&648 & \mc{\nodata} &   8&8 &  0&015 & \mc{\nodata} &  0&035 &  0&015&    622 &   131 \\
\hline
         0&1 &   1&52 &         0&10 &         2&35 &         2&29 &    200 &   3&426 &        0&239 &  47&5 &  0&079 &        0&351 &  0&170 &  0&061&    127 &    20 \\
\mc{\nodata} &   1&24 & \mc{\nodata} & \mc{\nodata} & \mc{\nodata} &    300 &   2&283 & \mc{\nodata} &  47&0 &  0&053 & \mc{\nodata} &  0&114 &  0&041&    191 &    30 \\
\mc{\nodata} &   1&08 & \mc{\nodata} & \mc{\nodata} & \mc{\nodata} &    400 &   1&713 & \mc{\nodata} &  47&6 &  0&040 & \mc{\nodata} &  0&085 &  0&030&    254 &    40 \\
\mc{\nodata} &   0&96 & \mc{\nodata} & \mc{\nodata} & \mc{\nodata} &    500 &   1&370 & \mc{\nodata} &  47&4 &  0&032 & \mc{\nodata} &  0&068 &  0&024&    318 &    50 \\
\mc{\nodata} &   0&88 & \mc{\nodata} & \mc{\nodata} & \mc{\nodata} &    600 &   1&142 & \mc{\nodata} &  47&5 &  0&026 & \mc{\nodata} &  0&057 &  0&020&    382 &    61 \\
\mc{\nodata} &   0&81 & \mc{\nodata} & \mc{\nodata} & \mc{\nodata} &    700 &   0&979 & \mc{\nodata} &  47&7 &  0&023 & \mc{\nodata} &  0&049 &  0&017&    446 &    71 \\
\mc{\nodata} &   0&76 & \mc{\nodata} & \mc{\nodata} & \mc{\nodata} &    800 &   0&856 & \mc{\nodata} &  47&5 &  0&020 & \mc{\nodata} &  0&043 &  0&015&    509 &    81 \\
\mc{\nodata} &   0&68 & \mc{\nodata} & \mc{\nodata} & \mc{\nodata} &   1000 &   0&685 & \mc{\nodata} &  47&2 &  0&016 & \mc{\nodata} &  0&034 &  0&012&    637 &   101 \\
\hline
         1&0 &  14&89 &         0&01 &         6&54 &         0&50 &    200 &   5&026 &        0&501 &   5&8 &  0&047 &        0&508 &  0&365 &  0&337&     76 &    33 \\
\mc{\nodata} &  12&16 & \mc{\nodata} & \mc{\nodata} & \mc{\nodata} &    300 &   3&350 & \mc{\nodata} &   6&7 &  0&031 & \mc{\nodata} &  0&244 &  0&224&    114 &    49 \\
\mc{\nodata} &  10&53 & \mc{\nodata} & \mc{\nodata} & \mc{\nodata} &    400 &   2&513 & \mc{\nodata} &   5&8 &  0&024 & \mc{\nodata} &  0&183 &  0&168&    152 &    66 \\
\mc{\nodata} &   9&42 & \mc{\nodata} & \mc{\nodata} & \mc{\nodata} &    500 &   2&011 & \mc{\nodata} &   8&9 &  0&019 & \mc{\nodata} &  0&146 &  0&135&    190 &    82 \\
\mc{\nodata} &   8&60 & \mc{\nodata} & \mc{\nodata} & \mc{\nodata} &    600 &   1&677 & \mc{\nodata} &   5&8 &  0&016 & \mc{\nodata} &  0&122 &  0&112&    228 &    99 \\
\mc{\nodata} &   7&96 & \mc{\nodata} & \mc{\nodata} & \mc{\nodata} &    700 &   1&437 & \mc{\nodata} &   6&0 &  0&013 & \mc{\nodata} &  0&104 &  0&096&    266 &   116 \\
\mc{\nodata} &   7&45 & \mc{\nodata} & \mc{\nodata} & \mc{\nodata} &    800 &   1&257 & \mc{\nodata} &  30&3 &  0&012 & \mc{\nodata} &  0&091 &  0&084&    304 &   132 \\
\mc{\nodata} &   6&66 & \mc{\nodata} & \mc{\nodata} & \mc{\nodata} &   1000 &   1&006 & \mc{\nodata} &   5&6 &  0&009 & \mc{\nodata} &  0&073 &  0&067&    381 &   165 \\
\hline
         0&1 &   4&50 &         0&01 &         7&30 &         2&00 &    200 &   4&893 &        0&341 &  11&8 &  0&055 &        0&492 &  0&298 &  0&227&     81 &    24 \\
\mc{\nodata} &   3&68 & \mc{\nodata} & \mc{\nodata} & \mc{\nodata} &    300 &   3&262 & \mc{\nodata} &  11&5 &  0&037 & \mc{\nodata} &  0&199 &  0&151&    122 &    36 \\
\mc{\nodata} &   3&18 & \mc{\nodata} & \mc{\nodata} & \mc{\nodata} &    400 &   2&446 & \mc{\nodata} &  11&2 &  0&028 & \mc{\nodata} &  0&149 &  0&113&    163 &    48 \\
\mc{\nodata} &   2&85 & \mc{\nodata} & \mc{\nodata} & \mc{\nodata} &    500 &   1&957 & \mc{\nodata} &  11&8 &  0&022 & \mc{\nodata} &  0&119 &  0&091&    204 &    60 \\
\mc{\nodata} &   2&60 & \mc{\nodata} & \mc{\nodata} & \mc{\nodata} &    600 &   1&631 & \mc{\nodata} &  11&5 &  0&018 & \mc{\nodata} &  0&099 &  0&076&    244 &    72 \\
\mc{\nodata} &   2&41 & \mc{\nodata} & \mc{\nodata} & \mc{\nodata} &    700 &   1&398 & \mc{\nodata} &  12&6 &  0&016 & \mc{\nodata} &  0&085 &  0&065&    285 &    84 \\
\mc{\nodata} &   2&25 & \mc{\nodata} & \mc{\nodata} & \mc{\nodata} &    800 &   1&223 & \mc{\nodata} &  12&5 &  0&014 & \mc{\nodata} &  0&075 &  0&057&    326 &    96 \\
\mc{\nodata} &   2&01 & \mc{\nodata} & \mc{\nodata} & \mc{\nodata} &   1000 &   0&979 & \mc{\nodata} &  11&2 &  0&011 & \mc{\nodata} &  0&060 &  0&045&    408 &   120 \\
\enddata

\tablecomments{{}Parameters of models. (1): External magnetic field. (2): Final mean magnetic field (eq.~\ref{eq:b}). (3): Inital pressure. (4) \& (5): Final, volume-averaged Mach numbers. (6): Box size. (7) \& (8): Most probable \EM\ and standard deviation of $\log \EM$. These data with a box size of $500 \pc$ are plotted in Figure~\ref{fig:ms_ma_em}. (9): Reduced $\chi^2$ goodness-of-fit parameter of lognormal fit to simulated EM distribution. (10) \& (11): Most probable density and standard deviation of $\log n$. (12) -- (15): Characteristic density and occupation length (see \S~\ref{sec:f}).}
\label{tbl:ms_ma_em}

\end{\deluxetablestar}

\notetoeditor{If Table~\ref{tbl:ms_ma_em} does not fit on one page, it should be formatted so that any page breaks lie between groups of related parameters (i.\ e.\ at {\tt hline} commands, not splitting rows with {\tt nodata} commands in some columns). If the table fits on the page vertically, that would be preferable to the horizontal layout necessary in the preprint format.}

Each of the twelve KLB models used here is uniquely identified by two parameters: the initial thermal pressure, \Pini, and a uniform external magnetic field strength, \Bext. In dimensionless units, the values used were $\Bext=0.1$ (``weakly'' magnetized) and $1.0$ (``strongly'' magnetized), and $\Pini=0.01$, $0.1$, $0.2$, $1.0$, $4.0$, and $8.0$. After a simulation has reached a steady-state, it can be characterized by a final mean sonic Mach number \Ms\ and mean Alfv\'en Mach number \Ma. The sonic Mach number decreases with increasing initial pressure; as \Pini\ rises from $0.01$ to $1.0$, \Ms\ drops from $7$ to $0.7$. The Alfv\'en Mach number rises with decreasing \Bext, yielding $\Ma=0.5-0.6$ for high \Bext\ cases, and $\Ma=2.0-2.3$ for low \Bext\ cases. We note that these Mach numbers are {\it volume} averages. This may overemphasize the impact of low density cells which do not contribute strongly to the observed properties of the gas but have systematically higher velocities than more readily-observed, higher density cells. However, the mean Mach number of the densest $5 \%$ of the cells in the simulation cube typically differs from the mean over the entire cube by less than $10 \%$.

Using the physical scaling discussed above, we calculate simulated emission measures for each model (Figure~\ref{fig:kowal_em}) to determine which model, if any, provides the best match to the WHAM data. In each case, the simulated emission measure is calculated along each of $256^{2}$ sightlines oriented perpendicular to the external magnetic field. (For the sonic Mach number required to match the data, the predicted \emsinb\ distribution does not depend significantly on the orientation with respect to the magnetic field, as discussed in \S~\ref{sec:magnetic_field}.) Parameters characterizing the \emsinb\ and density distribution for each model are given in Table~\ref{tbl:ms_ma_em}.

Figure~\ref{fig:ms_ma_em} shows the mean and standard deviation of the lognormal distribution found in the WHAM observations compared to Gaussian fits to the lognormal distribution found in each of the models. A successful model must match both the mean, $\langle \log \emsinb \rangle$, and the standard deviation, $\sigma_{\log \emsinb}$, of the WHAM distribution of $\log \emsinb$. The standard deviation is highly sensitive to the sonic Mach number; higher and wider distributions of \EM\ correspond to higher sonic Mach numbers. Mildly supersonic ($1 \lesssim \Ms \lesssim 3$) simulations best match the width of the WHAM distribution. The spread in the \emsinb\ distribution is weakly sensitive to Alfv\'en Mach number; models with comparable pressure but \Ma\ varying by a factor of four have comparable \EM\ distributions. Note that the mean magnetic field strength in the strongly magnetized, $\Ms \sim 1-3$ models (column~2 of Tbl.~\ref{tbl:ms_ma_em}) is $2.4-3.4 \, \mu \mathrm{G}$; the WIM is thought to have a magnetic field strength of a few $\mu \mathrm{G}$ \citep{f01}.

\begin{figure}
\plotone{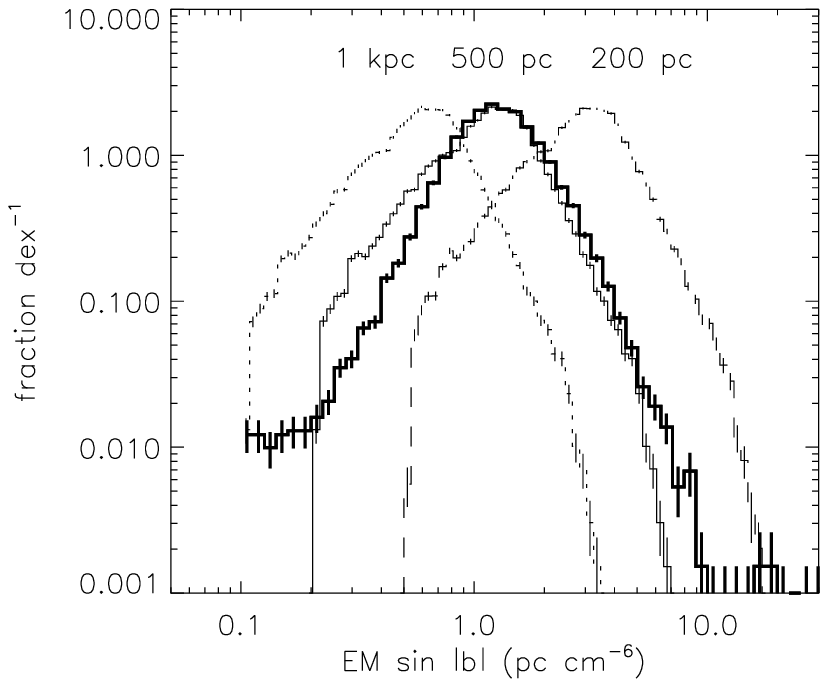}
\caption{Effects of changing the simulation box size. The WHAM WIM sightline histogram is shown with a solid, thick line. Three scalings of the $\Ms=1.7$ model are shown ({\em thin lines}) with box sizes of $200 \pc$ ({\em dashed}), $500 \pc$ ({\em solid}), and $1000 \pc$ ({\em dotted}).}
\label{fig:box_size}
\end{figure}

Once an appropriate value of \Ms\ is selected to match the width of the $\log \emsinb$ distribution, the physical box size can be varied to match the mean value.   For a fixed box size, high \Ms\ models with a relatively wide distribution of \EM\ have higher mean values of \EM\ than low \Ms\ models. Because \EM\ is the integral of $n^2$, the minimum value of \EM\ for a given column density ($\int n \, ds$) is for a perfectly homogeneous medium, whereas the maximum \EM\ for a given column density occurs when the gas is heavily clumped. Figure~\ref{fig:box_size} shows the effect of changing the box size of the model which best matches the observed width of the distribution. A box size of $\hsimbox \sim 400 - 500 \pc$ best matches the mean \EM\ for the mildly supersonic models. For the remainder of this paper, we adopt a box size of $500 \pc$.

\subsection{Effects of viewing geometry} \label{sec:visualization}

\begin{figure*}
\plotone{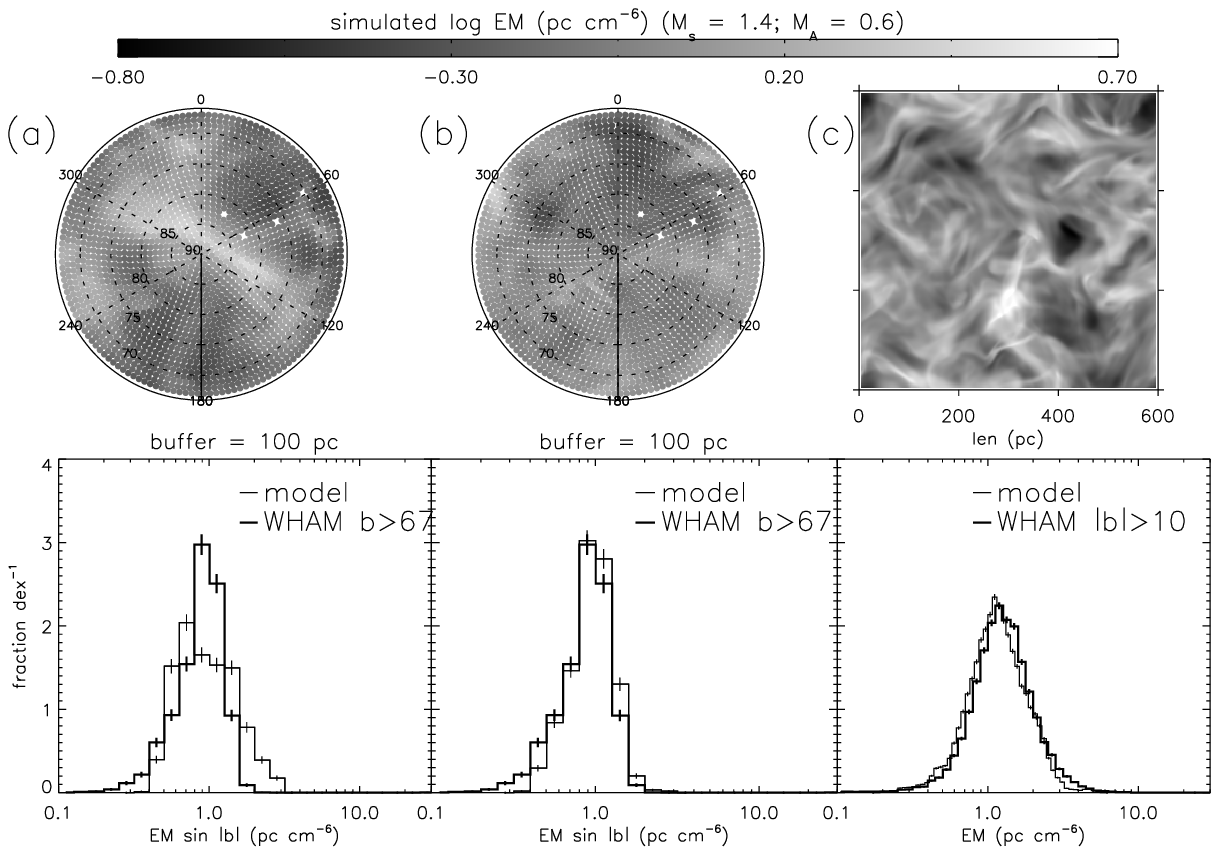}
\caption{Three visualizations of a strongly magnetized ($\Ma = 0.6$), mildly supersonic ($\Ms = 2.1$) model. In panel ($c$), emission measures are shown from the vantage point of each point along one surface of the cube looking vertically in parallel sightlines. In panel ($a$), $\EM \sin |b|$ is shown in all directions from the vantage point of the center of that plane with a $100 \pc$-thick plane with density $n=0$ below the cube. (Sightlines below $b = 67 \arcdeg$, which do not include the opposite face of the cube, are not shown.) Panel ($b$) is similar to panel ($a$), but the box has been shifted by half its length in one direction, exploiting the periodic boundary conditions. Corresponding histograms are shown below each panel.}
\label{fig:sim_maps}
\end{figure*}

In the previous section, we implicitly  placed an observer at different points along the plane at one face of the cube to look vertically through the cube with parallel sightlines. This visualization, used by KLB, has the advantage of simplicity and efficiently utilizes every cell of the simulation cube. However, in reality, we only enjoy one vantage point, which could potentially be biased by nearby features in the WIM. We thus consider the effects of a different viewing geometry. 

Here, we place the observer at the center of one face of the cube and look through the cube with diverging sightlines above a ``latitude'' of $65 \arcdeg$. Sightlines below $63 \arcdeg$ emerge from the side of the cube before sampling its full height. This visualization more accurately reflects observations of the WIM but is sensitive to the details of the vertical distribution of the gas, which is poorly understood and is not modelled by these simulations. Regions with a relatively small volume and an enhanced or depressed density that are near the observer cover a large solid angle and can therefore skew the distribution of intensities in this visualization. However, due to the Local Bubble, there is little WIM plasma within $\sim 100 \pc$ of the Sun; placing a gap of $100 \pc$ between the observer and the simulation cube lessens the impact of a small, nearby clump and yields distributions roughly comparable to the observed WIM distribution.

Samples of these visualizations are shown in Figure~\ref{fig:sim_maps}. Diverging sightline visualizations yield emission measure distributions consistent with those in the parallel sightline visualization. Because the diverging sightline visualization of the models is limited to $|b| > 65 \arcdeg$ and requires detailed knowledge of the vertical distribution of the gas to be realistic and because there is no evidence that adopting this geometry significantly changes the results, we use the parallel sightline visualization for the remainder of this paper.

\section{Testing additional predictions of isothermal MHD turbulence models} \label{sec:tests}

Although the KLB models of isothermal turbulence were not designed for the WIM, we find that they match WHAM \emsinb\ data for physically reasonable values, namely sonic Mach numbers in the range $\Ms = 1.4-2.4$ with a box length of $\hsimbox=500$ pc. With the exception of the magnetic field strength, which only weakly influences the \emsinb\ distribution, there are no additional parameters which are varied in the models used in this work.  We now present the results of testing model predictions of pulsar dispersion measures and \Ha\ velocity profiles, which provide additional support for this model. 

\subsection{DM distribution} \label{sec:dm}

\begin{figure*}
\plotone{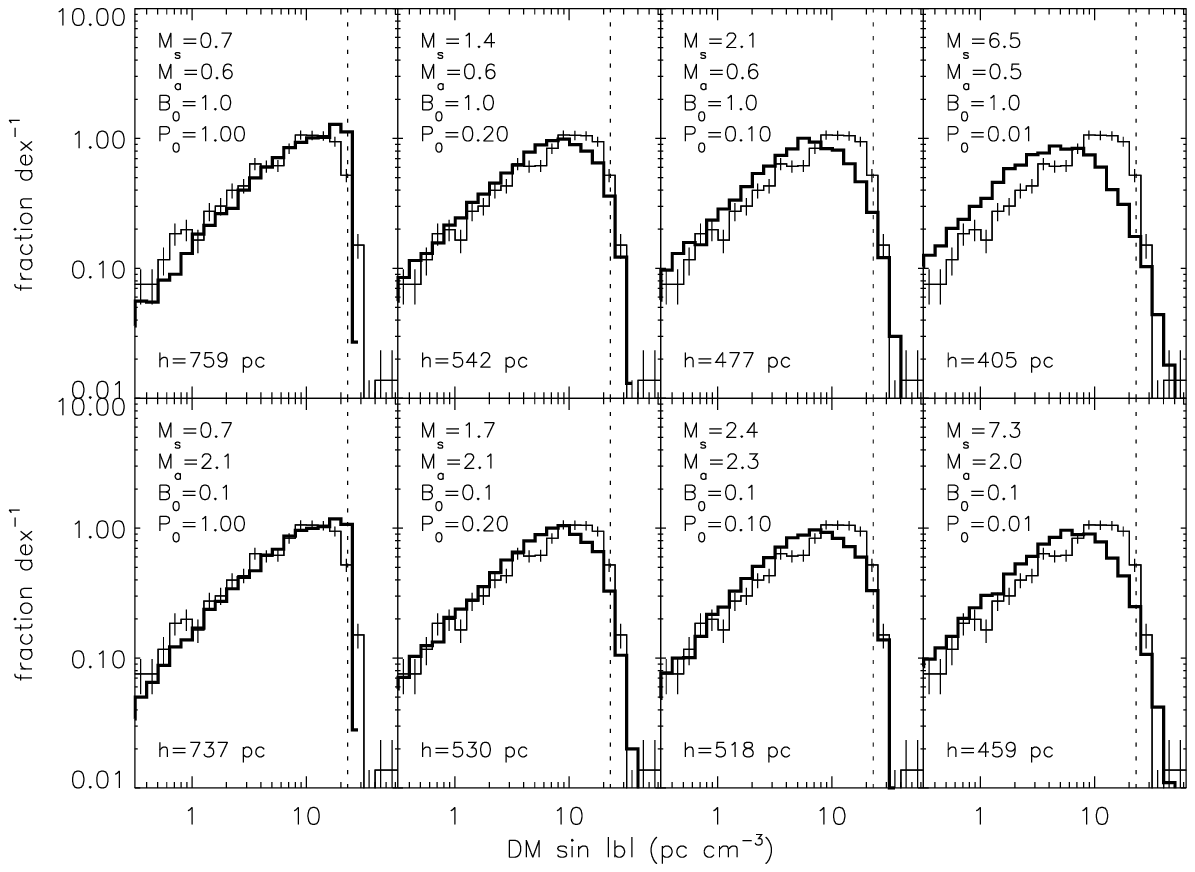}
\caption{Histograms of pulsar dispersion measures $\DM \sin |b|$, or column densities from the midplane to the pulsar ({\em thin lines}). The mean $\DM \sin |b|$ of high-$|z|$ globular clusters, which was used to define the mean column density through the simulations, is shown with a dashed line. The results of placing ``pulsars'' at random heights within the MHD simulations with $\hsimbox=500 \pc$ and the indicated pulsar scale heights ($h$) are also shown ({\em thick lines;} see \S~\ref{sec:dm}).}
\label{fig:kowal_dm}
\end{figure*}

Pulsar dispersion measures probe the column density of free electrons, $\DM = \int_0^D n_e ds$, where $D$ is the distance to the pulsar (which is known independently of \DM\ for relatively few pulsars). Figure~\ref{fig:kowal_dm} shows the model \DM\ distributions and the distribution for all pulsars with measured dispersion measures in the ATNF Pulsar Catalog \citep{mht05}, obtained on 2006 May 26 from the pulsar catalog web site.\footnote{http://www.atnf.csiro.au/research/pulsar/psrcat/} Because pulsars are embedded in the medium, \DM\ is a truncated column density, while emission measure probes the medium to infinity. The scale height of pulsars is poorly known, but may be comparable to the $1 \kpc$ scale height of the WIM \citep{l95}. If there were a sufficiently large population of pulsars with $|z| > 1 \kpc$, one would expect to see a pile-up of pulsars at the asymptotic value of $\DM \sin |b|=23 \pc \cucm$ measured with the $|z| > 3 \kpc$ globular cluster pulsars.  The spread in the asymptotic value would be due to the fluctuations in the dispersion measure. For the models considered, the predicted total column density is reasonably well fit by a lognormal distribution with $\sigma_{\log \DM, \mathrm{model}}=0.03-0.19$ over the full parameter space. For our preferred models with $\Ms=1.4$ to $2.4$, we find $\sigma_{\log \DM, \mathrm{model}}=0.08-0.12$.  By comparison, the six globular clusters pulsars are characterized by $\sigma_{\log \DM \sin |b|, \mathrm{obs}}=0.08$.

To test the dispersion measure predictions of the model, we randomly placed $10000$ pulsars within a simulated WIM. We assumed that the pulsars have a $\textrm{sech}^2(z)$ distribution (chosen because, unlike an exponential distribution, it does not have a cusp at $z=0$) and that the WIM has a uniform filling fraction up to $|z|=1 \kpc$. We then placed each pulsar in a random cell in the (two-dimensional) modelled column density grid and multipled the total column density in the cell by $z_\mathrm{pulsar} / \hwim$ or by one if $z_\mathrm{pulsar} > \hwim$. Of the 1428 non-globular cluster, Galactic pulsars (at all latitudes), $2.4 \%$ have $\DM \sin |b| > 23 \pc \cucm$. For each model, we chose the pulsar scale height so that the minimum height for which the fraction of sightlines with $\DM \sin |b| > 23 \pc \cucm$ matches this observed value. The required pulsar scale height (listed in Figure~\ref{fig:kowal_dm}) decreases from $\sim 750 \pc$ at $\Ms \approx 0.7$ to $\sim 450 \pc$ at $\Ms \approx 7$ and is $\sim 500 \pc$ in our preferred, mildly supersonic models. The required scale height is slightly higher in strongly magnetized simulations than in weakly magnetized ones.

The resulting distribution of vertical dispersion measures is shown in Figure~\ref{fig:kowal_dm} with a thick line. In all models, the observed low \DM\ tail and turnover near the observed peak at $\DM = 23 \pc \cucm$ are both reasonably well matched by the model.
Higher \Ms\ models under-predict the number of $\DM \sin |b| \approx 20 \pc \cucm$ pulsars and over-predict the number of $\DM \sin |b| \approx 2 \pc \cucm$ pulsars. The subsonic and preferred, mildly supersonic simulations each produce a DM distribution similar to the observed distribution of real pulsars.

In principle, because we constrain the sonic Mach number of the WIM independently of the DM distribution, these models allow a measurement of the pulsar scale height. However, the crude estimate of the $\DM \sin |b|$ distribution we present here does not consider any selection effects in the distribution of known pulsars with distance or latitude or multiple disk populations of pulsars. Moreover, the resulting pulsar scale height (or heights) would likely be substantially different if we employed a stratified-density model of the WIM.

\subsection{Velocity profiles} \label{sec:lineprof}

\begin{figure}
\plotone{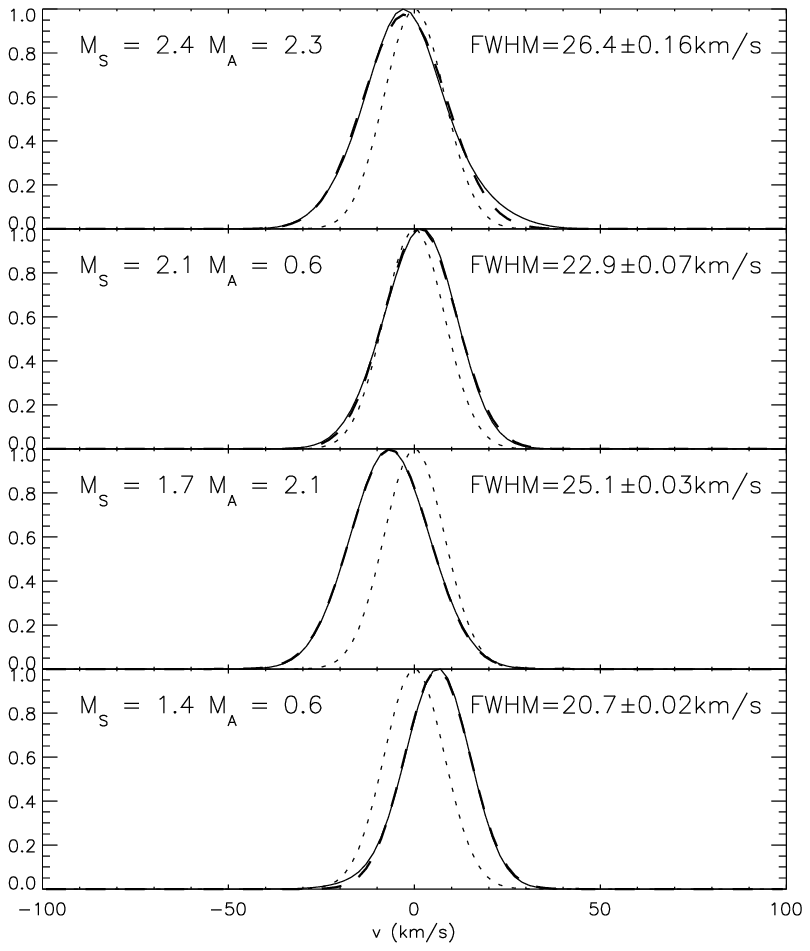}
\caption{Simulated line profiles of \Ha\ emission for models consistent with the observed \emsinb\ distribution, averaged over a $7$ cell square within the simulation cube. A single-component Gaussian fit is shown for each profile with a dashed line; the full width at half maximum of the Gaussian is listed. Dotted lines depict the thermal line profile at $8000 \K$ (centered at $v=0 \kms$) used to calculate the simulated profiles (see \S~\ref{sec:lineprof}). Intensities are in normalized units.}
\label{fig:sim_lineprofs}
\end{figure}

\begin{figure}
\plotone{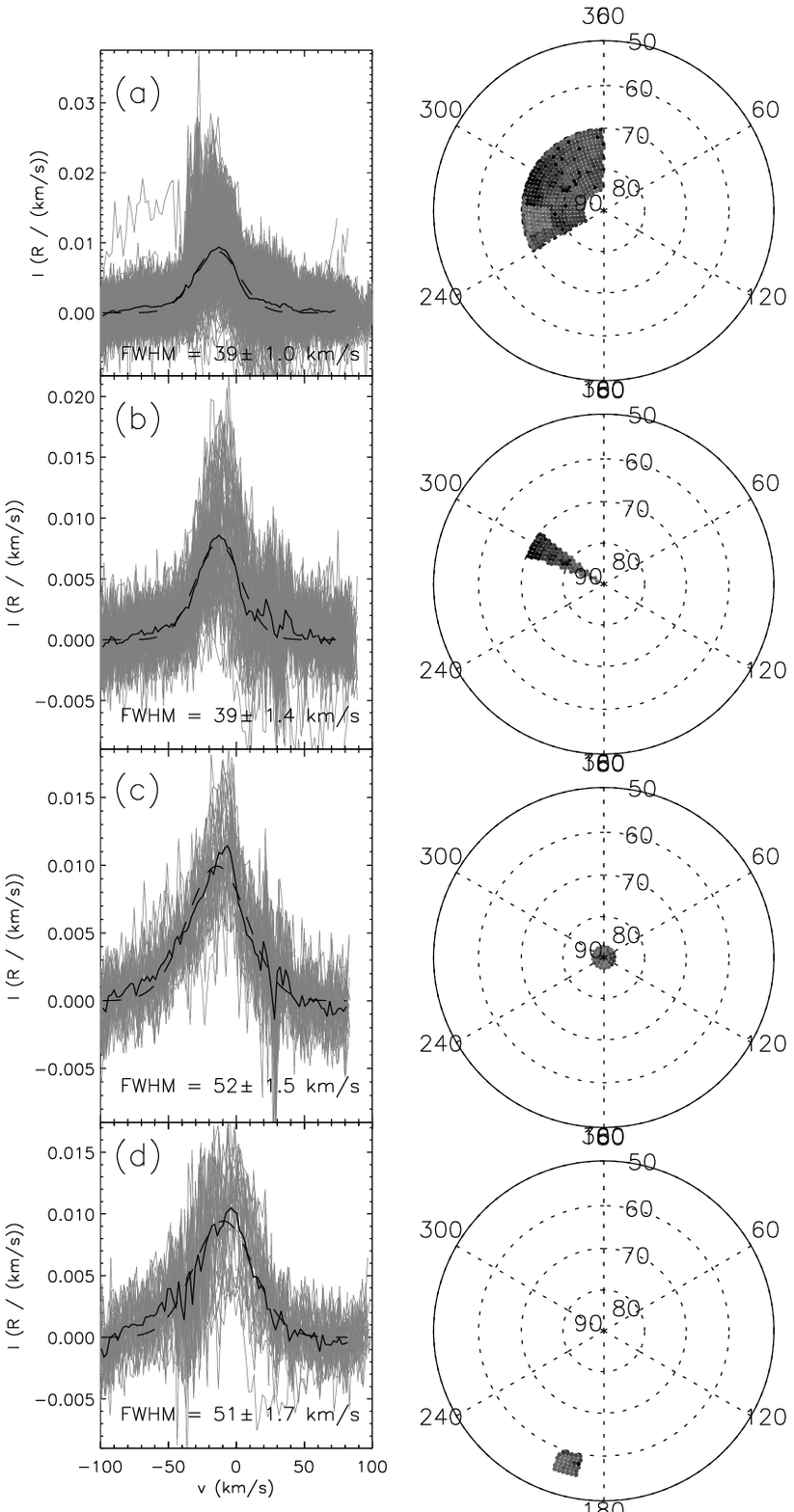}
\caption{\Ha\ line profiles from the WHAM survey. Each profile is an average of the line profiles (times $\sin |b|$) in the sightlines shown in the corresponding map. As in Fig.~\ref{fig:sim_lineprofs}, a single-component Gaussian fit is shown for each profile with a dashed line. Grey lines show the spectra used to compute the average line profile. Maps of the regions from which spectra were taken are shown at the right. Panel $(d)$ is the region known as the Lockman Window.}
\label{fig:lineprofs}
\end{figure}

To determine simulated velocity profiles, we assumed a thermal velocity profile in each cell with a center corresponding to the component of the simulated velocity along the line of sight. Along a line of sight at coordinates $(x, y)$ at a given velocity $v$, the simulated line intensity is
\begin{equation}
\EM(v) = \int n(\vec{x})^2 f(v, \vec{x}) dz = \sum_i n(x, y, z_i)^2 f(v, x, y, z_i),
\end{equation}
where $z$ is one of the coordinates represented by the position vector $\vec{x} \equiv (x, y, z)$. In a cell at a position $\vec{x}$ with velocity $v_0$, the thermal velocity profile is \citep[eq. 3-20 of][]{s78}
\begin{equation}
f(v, \vec{x}) = \frac{e^{- \left(v-v_0(\vec{x})\right)^2/b^2}}{b\sqrt{\pi}},
\end{equation}
where $b^2 = 2 k_B T / m$ is the square of the velocity width parameter; at $8000 \K$, $b = 11.5 \kms$ for atomic hydrogen, corresponding to a line full width at half-maximum of $19.1 \kms$. A $1 \arcdeg$ beam covers approximately $17 \pc$ at a distance of $1 \kpc$, and the size of an individual cell in the models is $2.3 \pc$, so we average over parallel sightlines in a 7~cell square in the simulation cubes to approximately match a single WHAM beam.
We show sample simulated line profiles for the mildly supersonic models in Figure~\ref{fig:sim_lineprofs}. Gaussian fits to many line profiles yield line widths varying from $20-25 \kms$ for $\Ms \approx 2.4$ models and $18-21 \kms$ for $\Ms \approx 1.4$ models.

How do these values compare to observational constraints?  WHAM survey \Ha\ line profiles near the Galactic pole or towards the Galactic anticenter ($l = 180 \arcdeg$) diagnose the motions of the gas independent of differential rotation. The polar regions have coherent regions of rising and falling gas (Fig.~14 of \citealt{hrt03}); positive velocity (rising) sightlines tend to have lower intensities than the predominant negative velocity (falling) sightlines. In Figure~\ref{fig:lineprofs}, we show average line profiles for four regions. In computing the line profiles, we multiplied the intensity in each pointing by $\sin |b|$ to correct for path length through the WIM. For these directions, \Ha\ line profiles have a full width at half maximum of $40 - 50 \kms$. This is approximately twice as wide as the simulated profiles. However, downward moving intermediate velocity gas contributes significantly to the observed line profiles, as well as poorly understood regions of large-scale bulk flows.  (The largest scale turbulent motions would be larger than the box size for our simulations, and therefore not predicted by our current model.)

Perhaps a better test of predictions of the model lies in choosing lines of sight that are known to be free of such large-scale motions.  \citet{hrht02} obtained \Ha\ line profiles towards two high-latitude sightlines with long WHAM integrations. With $9600$~s of integration (compared to $30$~s integrations used in the WHAM survey) towards the O star \object{HD 93521} $(l = 183.1 \arcdeg, b = +62.2 \arcdeg)$, they identified a local gas component centered at $\vlsr = -10 \pm 4 \kms$ with a FWHM of $22 \pm 6 \kms$, where $\vlsr$ is the velocity with respect to the local standard of rest. An additional, intermediate velocity component centered at $\vlsr = -51 \pm 3 \kms$ has a width of $39 \pm 7 \kms$. The sensitive observations allow the identification of separate components which are not typically resolved in the lower signal-to-noise survey exposures. If the HD~93521 sightline is representative and the low-velocity component corresponds to the WIM gas, the observed line profiles are therefore comparable in width to the simulated profiles. We note that if model line profiles had been {\it wider} than observations, the model would have been definitively ruled out. 

\section{Implications for the warm ionized medium} \label{sec:implications}

\subsection{The meaning of ``filling fraction''} \label{sec:f}

\begin{figure}
\plotone{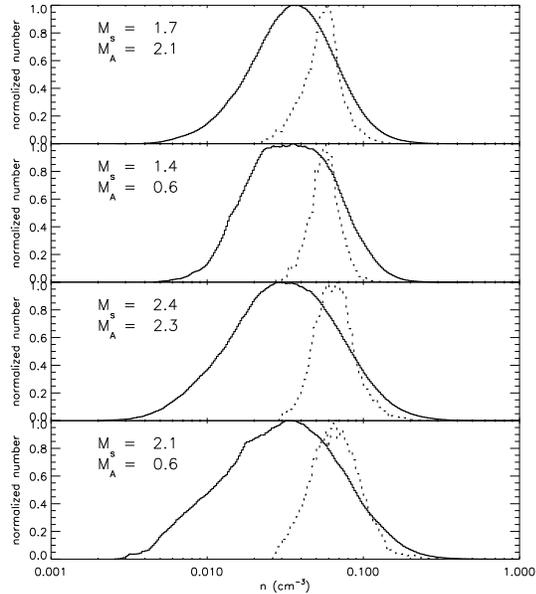}
\caption{Histograms of density ({\em solid line}) and characteristic density $n_c = \EM / \DM$ ({\em dotted line}) for mildly supersonic models.}
\label{fig:n_nc}
\end{figure}

Traditionally, density variations in the WIM have been characterized by an ``occupation length''
\begin{equation}
L_c = \frac{\DM^2}{\EM}
\end{equation}
and ``characteristic density''
\begin{equation}
n_c = \frac{\EM}{\DM}
\end{equation}
of discrete clumps of some fixed density. These parameters assume a bimodal density distribution function in which the density is either $n_c$ or zero, with clumps of density $n_c$ occupying a length $L_c$ along the line of sight. The filling fraction will depend upon the $n(z)$ distribution. For a uniform distribution, the filling fraction is $f = L_c/\hwim$; for a plane-parallel, exponential distribution, $f = L_c / 2\hwim$ \citep{r91}.

These values for the WIM can be obtained by combining \Ha\ emission measures and pulsar dispersion measures along lines of sight towards pulsars in high-latitude globular clusters more than a few scale heights above the midplane. For \object{M13}, \object{M15}, \object{M30}, and \object{M53}, we used the WHAM survey beam which includes the cluster to determine the EM. Sightlines towards \object{M3} and \object{M5} are contaminated by bright stars (\object{SAO 82944} and \object{SAO 120946}, respectively, which each have magnitudes $V < 7$); Fraunhofer lines from the stars cause an underestimate of the emission measure along those sightlines in the WHAM survey. Therefore, we used an average of sightlines with beam centers $\le 1.5 \arcdeg$ from the globular clusters, excluding the central beam, to estimate the EM. For the six globular clusters, the mean values are $\langle n_c \rangle \approx 0.07 \cucm$ with a filling fraction of $\approx 30 \%$ of a uniform density WIM disk ($14 \%$ of an exponential disk) with a scale height of $1 \kpc$, consistent with the results of \citet{r91}.

Consideration of a lognormal density distribution Êmodifies this physical interpretation. The filling fraction within the simulation cube is one, but the density within the cube has a lognormal distribution of values rather than a single value. The total filling fraction of the WIM within the Galaxy is now the volume which the simulation cube must occupy in order to reproduce the observed emission measure distribution divided by the total volume of the Galaxy within the layer in which the WIM resides. In place of the occupation length, $L_c$, Êand characteristic density, $n_c$, we have the simulation length, $\hsimbox$, and a distribution of densities characterized by the most probable density, $n_{pk} \equiv 10^{\langle \log n \rangle}$, and the standard deviation, $\sigma_{\log n}$ (columns~10 and 11 of Tbl.~\ref{tbl:ms_ma_em}, respectively). Using the synthetic \EM\ and \DM\ values derived from the models, we also calculate synthetic $n_c$ and $L_c$ values. Over the mildly sub/super-sonic range, $\Ms=0.7$ to $2.5$, $n_{pk}/n_c$ ranges from 0.95 to 0.4, and $L_c/\hsimbox$ ranges from $0.95$ to $0.62$. (For the highly supersonic case, these ratios are $\sim$ $0.1$ and $0.4$, respectively.) These changes are easily understood. Figure~\ref{fig:n_nc} shows histograms of the characteristic density for $256^2$ independent lines of sight through the mildly supersonic models. The characteristic density is considerably higher than the most probable density; $n_c$ is weighted to higher densities because of the $n^2$ dependence of \EM.

The effect of the different density distribution is to increase the volume that one would say is occupied by the WIM. Using the best fit box size, $\hsimbox=500 \pc$, and the observed WIM scale height, $\hwim = 1000 \pc$, $50 \%$ of the volume within the $1 \kpc$ thick layer above the midplane is occupied by the WIM with a most likely density of Ê$0.03 \cucm$. If the true scale height is $\hwim=1.8 \kpc$ \citep{gmc08}, the filling fraction would be smaller but our study's requirement for the box size would not change. In an exponentially stratified model, we expect that the mean density at the midplane would be similar to that in the uniform density model used here, but the filling fraction would be approximately half this value.

In reality, the WIM consists of an unknown number of pockets embedded in an ISM with cold and warm neutral and hot ionized regions, rather than the single, isothermal box employed here. Although the WIM itself is observationally approximately isothermal (\S~\ref{sec:data}), interfaces and mixing with other phases may affect the gas in ways not modelled here. \citet{db04} model the Galactic ISM with supernova explosions energizing a four-phase medium with heating and radiative cooling and find a combined volume filling factor of $\sim 0.6$ for the ``cool'' ($10^3 < T < 10^4$) and ``warm'' ($10^4 < T < 10^{5.5}$) gas. They find that pressure equilibrium between the phases does not hold; instead, turbulent mixing leads to a dynamical equilibrium. There is also a large range of temperatures that result, including gas that should be thermally unstable. ÊHowever, this and similar works do not generally take into account the ionization state of the gas by tracking the collisional and photo-ionization, recombination and radiative transfer through the gas. Presumably the gas identified as ``cool'' and ``warm'' in these simulations are a mixture of warm neutral and warm ionized components. Thus it is difficult to know what the density distribution of the observationally selected warm ionized gas would be in such a model.

\subsection{Power requirement for a turbulent warm ionized medium} \label{sec:power}

All of the KLB simulations we have considered were driven with the same forcing function. After the simulations reach a statistical steady-state, the energy injection equals the energy ``radiated'' away in the simulation. For isothermal simulations, the energy loss rate is assumed to be narrowly localized in the region of shocks and compressions rather than determined by the balance of heating and cooling. All of the simulations were driven with a power (in dimensionless units) of unity.  As shown in the Appendix, this can be converted to a physical value.

For our best-fitting simulations ($\Ms =1.4-2.4$, with $\hsimbox=500 \pc$), the gas reaches a turbulent steady-state with a kinetic and magnetic energy density of $U_\mathrm{kin}=(1.5-3) \times 10^{-13}~{\rm ergs~cm^{-3}}$ and  $U_\mathrm{mag}=(0.8-1.5) \times 10^{-13}~{\rm ergs~cm^{-3}}$.  The power density necessary to provide this turbulence is $P=(1.8-5.2)\times 10^{-25}~{\rm ergs~s^{-1}~cm^{-3}}$. Multiplying this value by $\hsimbox$ yields the power per area per side of the plane, $(3-8) \times 10^{-4}~{\rm ergs~s^{-1}~cm^{-2}}$. This is approximately the same power provided by $10^{51}$ ergs of supernova energy per 50 years spread over a disk of radius of $15 \kpc$ ($3 \times 10^{-4}~{\rm ergs~s^{-1}~cm^{-2}}$), but perhaps a bit on the high side. However, given the uncertainties in the spatial distribution of supernova energy input and the amount of energy converted into kinetic energy, these estimates should be viewed cautiously. 

Given the limitation of the isothermal assumption in our models, do our values provide upper or lower limits on the true power requirement of the WIM? The answer to this is not clear. One parameter in our models that would affect the power requirement is the driving scale, which was a quarter of the simulation box size. Using a smaller driving scale would probably increase the power requirement to maintain the same level of turbulence. Replacing the isothermal assumption (which assumes gas radiates immediately after shocks) with both supernova and photoionization heating and radiative cooling could raise or lower the power requirement for the WIM depending upon the temperature and density structure of the resulting medium. For the purpose of this work, it is comforting that the power requirement is not excessive. 

\subsection{Magnetic field structure of the warm ionized medium} \label{sec:magnetic_field}

\begin{figure}
\plotone{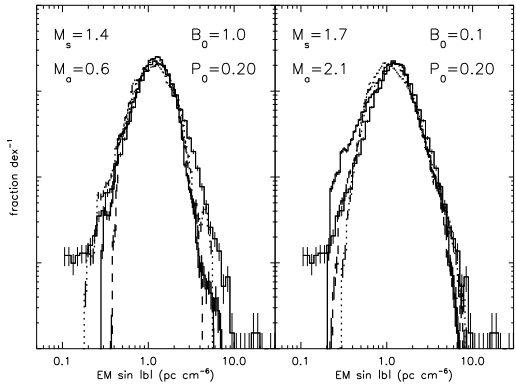}
\caption{Effect of orientation of line of sight with respect to the magnetic field. Thin lines show the observed, WHAM distribution. Thick lines show the modeled \EM\ distribution perpendicular to ({\em solid lines}) and parallel to the magnetic field ({\em dotted and dashed lines}).}
\label{fig:orientation}
\end{figure}

Figure~\ref{fig:orientation} shows \EM\ PDFs for each of the three possible orientations of the box (two perpendicular to the field and one along the field) for a weakly and strongly magnetized, mildly supersonic model. For each model, the three PDFs are similar, with counts slightly higher in the parallel view than in the perpendicular at low emission measures in the weakly magnetized model.

Because the distribution of density, column density, and emission measure in the models is relatively insensitive to the magnetic field strengths we considered, these data alone cannot effectively constrain the properties of the magnetic field. However, our study's density structure of the WIM effectively fixes the density, one of the free parameters in interpreting Faraday rotation studies of the ionized medium, which depends on the convolution of both the density and the magnetic field structure of the ISM at high Galactic latitude \citep{bc90, zh97, f01, h06, dkh06}.  Comparison of MHD models such as those we considered to the combination of density and magnetic field probes will be a powerful way to study the magnetic field structure of the WIM in the local solar neighborhood. 

We have checked to make sure that the magnetic fields used in our models included reasonable magnetic field stengths after normalizing the models to the data as described above. For the best fitting cases, the mean magnetic field strengths of the models considered are between $0.8-3.4 \mu \mathrm{G}$, similar to observational results \citep{f01}. A more extensive grid of magnetic field strengths would be desirable for comparison with Faraday rotation measure studies of high latitude extragalactic radio sources. 

\section{Summary and future directions}

Using data from the WHAM Northern Sky Survey with sightlines that intersect classical \ion{H}{2} regions removed, we find that the \Ha\ emission from the WIM has a lognormal distribution of \emsinb. Fits describing the observed distribution are listed in Table~\ref{tbl:emhist_fit}. If the WIM were a uniform, plane-parallel layer, \emsinb\ would be uniform at all latitudes, but the emission at $|b| > 60 \arcdeg$ is roughly $0.14$~dex lower than the emission in the range $10 < |b| < 30 \arcdeg$.

Because the WIM is an isothermal, magnetized plasma and a lognormal distribution arises from a series of compressions and rarefactions in a fluid, MHD models of isothermal turbulence appear to be applicable to the WIM. KLB recently completed such a simulation, which we apply to the physical scales and conditions of the WIM. We find that the model effectively reproduces the observed lognormal distribution of \Ha\ emission measures with mildly supersonic ($\Ms \sim 1.4-2.4$) turbulence, provided that the turbulent plasma fills roughly half of the $2 \kpc$-thick WIM layer. The model also successfully matches constraints from pulsar dispersion measures and \Ha\ line profiles. The model results in the WIM plasma filling roughly twice the fraction of the volume of the layer found in previous, relatively simplistic models \citep{r91}, with the most probable local density ($\approx 0.03 \cucm$) roughly half that found in previous models. We explore some of the ramifications of this result, including a discussion on the meaning of filling factor,  the power requirement of the turbulent WIM, the driving scale of turbulence, and implications for studies of the magnetized WIM. 

The KLB model used here was not designed specifically to describe the WIM; it is a generic model of isothermal turbulence. The model accurately reproduces the observations we have tested, but the solution is not unique and other models should certainly be tested against our observational result.  Probably the most important change to be considered in the future would be the development of vertically stratified models. Replacement of the isothermal assumption (appropriate for the WIM, which is observed to be nearly isothermal, but not the general ISM) with supernova and photoionization heating and radiative cooling may also be instructive.

\acknowledgments

We thank the referee, Enrique V\'{a}zquez-Semadeni, for a review which led to substantial improvements in the paper. WHAM is supported by the National Science Foundation through grants AST 0204973 and AST 0607512. G.\ K.\ and A.\ L.\ are supported by NSF grant AST 0307869 and the Center for Magnetic Self-Organization in Astrophysical and Laboratory Plasmas.

{\it Facility:} \facility{WHAM}

\appendix

\section{Physical scale of models} \label{app}

The simulations used in this work are conducted in dimensionless units. In order to compare observations directly with the simulations, we must impose reasonable physical constraints on the simulations. We adopt the notation that a tilde above a quantity denotes the quantity in dimensionless (simulation) units and a subscript zero denotes the scale factor; the symbol with neither a zero nor a tilde is the quantity in physical units. A modelled quantity in physical units is $a = a_0 \tilde{a}$. Observed constraints are denoted with the subscript `obs', and initial conditions that vary with the models are denoted with the subscript `ini'. 

The simulations used in this work output a density $\tilde{\rho}$ and $x$, $y$, and $z$ components of the magnetic field $\tilde{B}_{x, y, z}$ and velocity $\tilde{v}_{x, y, z}$ for each cell with grid coordinates $(\tilde{x}, \tilde{y}, \tilde{z})$. We specify, from observations, three physical scales: a length $x_0$, a velocity $v_0$, and a mass density $\rho_0$. These three scales are set by specifying the physical size of the simulation volume, the isothermal sound speed of the gas, and the dispersion measure of ionized gas through the simulation volume, respectively. 

\subsection{Length scale}

Since each grid cell is a cube, the volume of a cell is $1~{\rm unit}^{3}$; the  total volume of the simulation in dimensionless units is therefore $N_x \times N_y \times N_z$, where for the simulations considered here $N_x=N_y=N_z$. By specifying a physical length of the box, \hsimbox, we write the length scaling factor
\begin{equation}
x_0 = \frac{\hsimbox}{N_x}.
\end{equation}
Note that we treat the box size as a free parameter because the problem is overconstrained and the filling fraction of the WIM is not well known. The chosen box size must be smaller than the observed height of the WIM to be physically reasonable.

\subsection{Velocity scale}

The velocity scale factor comes from relating the sound speed in code units to that in physical units, 

\begin{equation}
v_0 = \frac{c_{s,\obs}}{\tilde{c_{s}}}=c_{s,\obs}\sqrt{\frac{\tilde{\rho}_\ini}{\tilde{p}_\ini}},
\end{equation}
where $p$ is the thermal pressure.
 
The physical sound speed of an ideal gas is
\[
c_s = \sqrt{\frac{\partial p}{\partial \rho}} = \sqrt{\frac{k_B T_\obs}{\mu}},
\]
where $T_\obs$ is the temperature and $\mu$ is the mean particle mass. The numerical value comes from  adopting $\bar{m} = 2.36 \times 10^{-24} \textrm{ g}$ as the mean mass per hydrogen atom, appropriate for a solar metallicity gas, and $\mu = \bar{m} / 2 y_e$, where $y_{e} \equiv n_{e}/n_{H}=1.1$ for a plasma in which the hydrogen is fully ionized and helium is singly ionized. This yields $c_s=10.15 \kms~(T / 8000~{\rm K})^{1/2}$.

For the simulation, the dimensionless sound speed of each model is
\[
\tilde{c}_s = \sqrt{\frac{\tilde{p}}{\tilde{\rho}}}.
\]
Initially, all cells are set to have a density $\tilde{\rho}_\ini = 1$; because there is no mass flow in or out of the simulation cube, the mean density remains unity at all times, although we leave it in the equations for dimensional clarity. The initial pressure $\tilde{p}_\ini$ is an input parameter for each model. The pressure is calculated in each cell by an isothermal equation of state, so $\tilde{p}/\tilde{\rho}$ is constant over time across the entire grid for a given simulation, and the sound speed is $\tilde{c}_s = (\tilde{p}_\ini/\tilde{\rho}_\ini)^{1/2}$ everywhere.

\subsection{Density scale}

The density scale factor comes from relating the surface mass density in the simulation to the surface mass density inferred from the observed dispersion measure and is given by 
\begin{equation}
\rho_0=\frac{\bar{m} \langle \DM \sin |b| \rangle_\obs}{y_{e}h_{\obs}\tilde{\rho}_\ini}.
\end{equation} 
Since each cell has a unit length in dimensionless coordinates, the dimensionless surface mass density through the simulation is 
\[
\tilde{\Sigma} = \sum_{i=0}^{N_x} \tilde{\rho}_i=\tilde{\rho}_\ini N_x. 
\]

The physical surface mass density can be derived by comparison to the mean vertical dispersion measure through the half-thickness of the warm ionized medium, where we use the dispersion measure $(\DM \sin |b|)_\obs$ of pulsars in globular clusters more than $3 \kpc$ above the midplane. This corresponds to a surface mass density of 

\[
\Sigma_\obs = \int_0^\infty \rho \, ds = \int_0^\infty n_H \bar{m} \, ds = \frac{\bar{m} \, (\DM \sin |b|)_\obs}{y_e}.
\]

We can relate the dimensionless and physical surface mass density to get the surface density scale factor, $\Sigma_{0}=\Sigma_\obs/\tilde{\Sigma}$. Using the fact that $\Sigma_{0}=\rho_{0}x_{0}$, we get the mass density scaling factor as above. 

\subsection{Additional scale factors}

The three scale factors above allow us to determine the values of the scale factor for other physical values including time, 
\begin{equation} 
t_{0}=\frac{x_{0}}{v_{0}}=\frac{h_\obs}{N_x c_{s,\obs}}\sqrt{\frac{\tilde{p}_\ini}{\tilde{\rho}_{ini}}}, 
\end{equation}
pressure and energy density (which have the same units),
\begin{equation} 
p_{0}=\rho_{0}{v_{0}}^2= \frac{\bar{m}  \, (\DM \sin |b|)_\obs c_{s,\obs}^2 }{  h_{\obs} y_{e}   \tilde{p}_\ini },
\end{equation} 
power density,
\begin{equation} 
P_{0}=\frac{p_{0}}{t_{0}}=\frac{\bar{m} \, (\DM \sin |b|)_\obs c_{s,\obs}^3 N_x }{ y_e h_{\obs} \tilde{\rho}_\ini \tilde{p}_\ini},
\end{equation} 
and magnetic field strength 
\begin{equation}
B_0 = \sqrt{\rho_0 v_0^2}= c_{s,\obs} \sqrt{ \frac{ \bar{m}  \, (\DM \sin |b|)_\obs }{ y_{e} h_{\obs} \tilde{p}_\ini } }.
\end{equation} 

Although we do not explore the effect of magnetic field strengths in this paper (because the density PDF is fairly insensitive to this parameter), we note that the mean magnetic field strength that develops in these simulations can be obtained by comparing the sonic and Alfv\'en Mach number recorded for each simulation and is given by 
\begin{equation} \label{eq:b}
\langle B \rangle = 0.519 \left( \frac{\Ms}{\Ma} \right) \left(\frac{n_{pk}}{0.01 \cucm}\right)^{1/2} \left(\frac{c_s}{10 \kms}\right) \, \mu \mathrm{G},
\end{equation}
where $n_{pk}$ is the most probable electron density. This field is a mixture of the originally imposed uniform component and the random component that developed as a result of the turbulent driving.

\bibliography{references}


\end{document}